\documentclass[traditabstract,a4paper]{aa}
\usepackage{amsmath}
\usepackage{graphicx}
\usepackage{color}
\usepackage{natbib}
\usepackage{amssymb}
\input{psfig}
\def\beq{\begin{equation}}
\def\eeq{\end{equation}}
\def\barray{\begin{eqnarray}}
\def\earray{\end{eqnarray}}
\def\beqarray{\begin{eqnarray}}
\def\eeqarray{\end{eqnarray}}

\def\rmc{{\rm c}}
\def\rmd{{\rm d}}

\def\rmg{{\rm g}}
\def\rmh{{\rm h}}

\def\rmm{{\rm m}}

\def\rmp{{\rm p}}

\def\rms{{\rm s}}

\def\rmx{{\rm x}}
\def\rmy{{\rm y}}

\def\calD{{\cal D}}

\def\calH{{\cal H}}

\def\calL{{\cal L}}

\def\bx{{\bf x}}

\def\bA{{\bf A}}

\def\bC{{\bf C}}

\def\gtsima{$\; \buildrel > \over \sim \;$}
\def\ltsima{$\; \buildrel < \over \sim \;$}
\def\prosima{$\; \buildrel \propto \over \sim \;$}
\def\gsim{\lower.7ex\hbox{\gtsima}}
\def\lsim{\lower.7ex\hbox{\ltsima}}
\def\simgt{\lower.7ex\hbox{\gtsima}}
\def\simlt{\lower.7ex\hbox{\ltsima}}
\def\simpr{\lower.7ex\hbox{\prosima}}

\newdimen\hssize
\newdimen\hdsize
\bibpunct{(}{)}{;}{a}{}{,}

\begin{document}

\author{Edo van Uitert \inst{\ref{inst1}}, Marcello Cacciato \inst{\ref{inst2}}, Henk Hoekstra \inst{\ref{inst2}} \and Ricardo Herbonnet \inst{\ref{inst2}} }
\institute{Argelander-Institut f\"ur Astronomie, Auf dem H\"ugel 71, 53121 Bonn, Germany \label{inst1} \and Leiden Observatory, Leiden University, Niels Bohrweg 2, NL-2333 CA Leiden, The Netherlands \label{inst2}}

\title{Evolution of the luminosity-to-halo mass relation of LRGs from a combined SDSS-DR10+RCS2 analysis}

\titlerunning{Luminosity-to-halo mass relation of LRGs}

\authorrunning{Van Uitert, Cacciato, Hoekstra \& Herbonnet}

\abstract {We study the evolution of the luminosity-to-halo mass relation of Luminous Red Galaxies (LRGs). We select a sample of 52 000 LOWZ and CMASS LRGs from the Baryon Oscillation Spectroscopic Survey (BOSS) SDSS-DR10 in the $\sim$450 deg$^2$ that overlaps with imaging data from the second Red-sequence Cluster Survey (RCS2), group them into bins of absolute magnitude and redshift and measure their weak lensing signals. The source redshift distribution has a median of 0.7, which allows us to study the lensing signal as a function of lens redshift. We interpret the lensing signal using a halo model, from which we obtain the halo masses as well as the normalisations of the mass-concentration relations. We find that the concentration of haloes that host LRGs is consistent with dark matter only simulations once we allow for miscentering or satellites in the modelling. The slope of the luminosity-to-halo mass relation has a typical value of 1.4 and does not change with redshift, but we do find evidence for a change in amplitude: the average halo mass of LOWZ galaxies increases by $25_{-14}^{+16}$\% between $z=0.36$ and 0.22 to an average value of $(6.43\pm0.52)\times 10^{13} h_{70}^{-1} M_\odot$. If we extend the redshift range using the CMASS galaxies and assume that they are the progenitors of the LOWZ sample, we find that the average mass of LRGs increases by $80^{+39}_{-28}\%$ between $z=0.6$ and 0.2.}


\maketitle


\section{Introduction}  
Hierarchical models of structure formation predict that galaxies form in small dark matter haloes, which subsequently clump together and merge into larger ones \citep{White78}. At large scales, the evolution of structure is mainly determined by the properties of dark matter and dark energy. However, at smaller, galactic scales, baryonic physics cannot be ignored. Processes such as supernova and AGN feedback impact the relation between the observable (baryonic) properties of galaxies and their dark matter haloes. Hence by measuring these relations, we gain insight into the processes that affected them. Studying this with numerical simulations is notoriously difficult, although in recent years this field has rapidly advanced through the use of semi-analytic models \citep[e.g.][]{Baugh06} and hydrodynamical simulations \citep[e.g.][]{Vogelsberger14,Schaye15}. To test these simulations and guide them with further input, we need observations of the relation between the properties of galaxies and their dark matter haloes. This is also crucial for understanding the effect of baryonic physics on the dark matter power spectrum \citep[e.g][]{VanDaalen11,Semboloni11}, which is the main observable in weak lensing studies that aim to extract cosmological parameters, such as Euclid \citep{Laureijs11}. \\
\indent The properties of dark matter haloes around galaxies can be studied with weak gravitational lensing. As the photons emitted by distant galaxies traverse the Universe, they are deflected due to the curvature of space around intervening mass inhomogeneities in the foreground. Consequently, the observed shapes of these background galaxies slightly deform, a distortion that can be reliably measured out to projected separations of tens of Mpcs around the lenses \citep[e.g.][]{Mandelbaum13}. Since this completely covers the regime where the dark matter halo of any lens dominates, weak gravitational lensing offers an excellent tool to measure halo masses. The weak lensing signal of individual galaxies is too noisy to be detected, but by averaging the signal of many lenses of similar observable properties, e.g. in a certain luminosity range, we can learn about the average halo properties of such lens samples. \\
\indent The relation between the properties of galaxies and their dark matter haloes has been studied before with weak lensing \citep[e.g.][]{Hoekstra05,Mandelbaum06,Li09,VanUitert11,Brimioulle13,Velander13}, but most of these studies focused on lenses at a limited redshift range. However, to study how galaxies evolve, one would like to measure how the luminosity-to-halo mass relation depends on lookback time. Recent imaging surveys such as the Canada-France-Hawaii Telescope Survey (CFHTLS) and the second Red-sequence Cluster Survey (RCS2) contain sufficient statistical power to enable such studies. Redshift dependent constraints that are derived in a homogeneous way, as is done in this study, are particularly useful for numerical simulations, as they can potentially disentangle degeneracies among the model parameters and limit the space for fine-tuning to match low-redshift observations \citep[for an example, see Figure 23 in][]{Guo11}. \\
\indent In this work, we study a particular type of galaxies: Luminous Red Galaxies (LRGs). They form an interesting subsample of the total population of galaxies, as they trace the highest density peaks in the Universe. These galaxies are thought to have formed around $z$$\sim$2 during a relatively short and intense period of star formation, after which the formation of stars practically halted. Their luminosity evolution can therefore be approximately described by `passive evolution', the evolution of a stellar population without forming new stars \citep[e.g.][]{Glazebrook04,Cimatti06,Roseboom06,Cool08,Banerji10}. This enables us to model the luminosity evolution, for example with stellar population synthesis models \citep[e.g.][]{Bruzual03,Conroy09,Conroy10,Maraston09}, and separate that from the halo mass evolution part. Low-level star formation and mergers may also contribute to the luminosity evolution of LRGs, but this is thought to mainly affect less massive LRGs \citep{Scarlata07,Pozzetti10,Tojeiro10,Tojeiro11,Tojeiro11m,Tojeiro12}. How large the average impact is on the luminosity evolution, compared to the pure passive evolution scenario, is not clear. However, for massive and luminous LRGs, the luminosity evolution is thought to be well understood.\\
\indent Also from an observational perspective, LRGs are advantageous to study. They are easily selected in multi-band optical data-sets and their redshifts can be relatively easily determined using the 4000{\AA} break \citep{Eisenstein01}. More than a million LRGs have been observed spectroscopically as part of the Baryon Oscillation Spectroscopic Survey \citep[BOSS;][]{Dawson13}, forming the LOWZ sample, which targets $z\lesssim0.4$ galaxies, and the CMASS sample, which targets $0.4<z<0.7$ galaxies. From a weak lensing perspective, the advantage of LRGs is that they are massive and therefore produce a large lensing signal that can be measured up to relatively high redshift. The overlap between the BOSS survey and the RCS2 therefore offers a perfect combined dataset to study the evolution of the luminosity-to-halo mass relation of LRGs. \\
\indent The outline of this work is as follows. In Section \ref{sec_data} we describe the data that we use in this work, how we compute the luminosities and how we perform the lensing analysis. We interpret the lensing measurements with the halo model, which we describe in Section \ref{sec_hm}. The evolution of the luminosity-to-halo mass relation is presented and discussed in Section \ref{sec_res}. The mass-concentration relation is discussed in Section \ref{sec_massconc}. We conclude in Section \ref{sec_conc}. Unless stated otherwise, we assume a WMAP7 cosmology \citep{Komatsu11} with $\sigma_8=0.8$, $\Omega_{\Lambda}=0.73$, $\Omega_{\rm M}=0.27$, $\Omega_{\rm b}=0.046$, and \mbox{$h_{70}=H_0/70$ km s$^{-1}$ Mpc$^{-1}$} with $H_0$ the Hubble constant; all distances are quoted in physical (rather than comoving) units; and all apparent magnitudes have been corrected using the dust maps from \citet{Schlegel98}.


\section{Data analysis \label{sec_data}}
In this work we make use of data from the tenth data release \citep[DR10;][]{Ahn14} from the Sloan Digital Sky Survey \citep[SDSS;][]{York00} and from the second Red-sequence Cluster Survey \citep[RCS2;][]{Gilbank10}. As in \citet{VanUitert11,VanUitert13,Cacciato14}, we take advantage that the SDSS contains more ancillary data on galaxies than is available in the RCS2 due to its photometry in five optical band and its spectroscopy for over a million of galaxies. However, the RCS2 imaging is $\sim$2 magnitudes deeper and achieved a median seeing of approximately $0.7''$, compared to $1.2''$ for SDSS, making the RCS2 better suited for a weak lensing analysis of lenses at higher redshifts. The total overlap between the RCS2 and the DR10 amounts to roughly 450 square degrees. \\
\indent A first combined analysis of the overlap between the ninth data release of SDSS \citep[DR9;][]{Ahn12} and the RCS2 was presented in \citet{Cacciato14}, where the lensing signal of the DR9 galaxies with spectroscopy was studied using RCS2 galaxies as sources. In that work we did not study the redshift evolution of the lensing signal, as, in contrary to the current work, we studied a mixed sample of early- and late-type galaxies, whose combined luminosity evolution was not well understood. Also, in DR9, the number of (high-redshift) BOSS spectra was roughly half of that in DR10. \\
\subsection{Lens sample}
\indent We use a subset of the total sample of overlapping DR10 galaxies with spectroscopy as our lenses, i.e. only the LRGs. We select all galaxies that have been targeted as part of BOSS. These are selected from the SDSS catalogues by requiring \footnote{\tt http://www.sdss3.org/dr9/algorithms/boss\_galaxy\_ts.php}
\begin{itemize}
\item[$\bullet$] BOSS\_TARGET1 \&\& 2$^0$
\item[$\bullet$] SPECPRIMARY == 1
\item[$\bullet$] ZWARNING\_NOQSO == 0
\item[$\bullet$] TILEID $>=$  10324
\end{itemize}
for the LOWZ sample, and
\begin{itemize}
\item[$\bullet$] BOSS\_TARGET1 \&\& 2$^1$
\item[$\bullet$] SPECPRIMARY == 1
\item[$\bullet$] ZWARNING\_NOQSO == 0
\item[$\bullet$] (CHUNK != ``boss1'') \&\& (CHUNK != ``boss2'')
\item[$\bullet$] $i_{\rm fib2}<21.5$
\end{itemize}
for the CMASS (high-z) sample. Additionally, we select all objects with reliable spectroscopy from the SDSS catalogues that satisfy the BOSS LOWZ target selection cuts:
\begin{itemize}
\item[$\bullet$] $|(r-i) - (g-r)/4-0.18|<0.2$ 
\item[$\bullet$] $r<13.5 + [0.7\times(g-r)+1.2\times((r-i)-0.18)]/3$
\item[$\bullet$] $16<\widetilde{r}<19.6$
\item[$\bullet$] SCIENCEPRIMARY==1
\item[$\bullet$] ZWARNING\_NOQSO == 0
\item[$\bullet$] $z_{\rm spec}>0.01$
\end{itemize}
where $g$, $r$ and $i$ indicate model magnitudes and $\widetilde{r}$ cmodel magnitudes. Note that we replaced the BOSS selection criterion \mbox{$r_{\rm psf} - r_{\rm cmod} > 0.3$} with $z_{\rm spec}>0.01$ to ensure that we have no stars. Finally, we also selected all objects that satisfied the CMASS selection cuts, but we found that all objects were already targeted and labeled as being BOSS galaxies, and it therefore did not increase the lens sample. \\
\indent Even though the LOWZ and CMASS samples mainly consist of LRGs, the populations differ due to the different colour and magnitude selection cuts. \citet{Tojeiro12} study which fraction of the CMASS LRGs are progenitors of the LOWZ sample, and find that this strongly depends on absolute magnitude, with the highest fractions found for the most luminous objects. A second but weaker trend is found with rest-frame colour. Therefore, we choose to analyse the LOWZ and the CMASS samples separately. We do investigate what we can conclude about the evolution of the luminosity-to-halo mass relation of LRGs if we consider the CMASS sample as progenitors of the LOWZ LRGs.
\begin{figure}[t!]
  \resizebox{\hsize}{!}{\includegraphics[angle=-90]{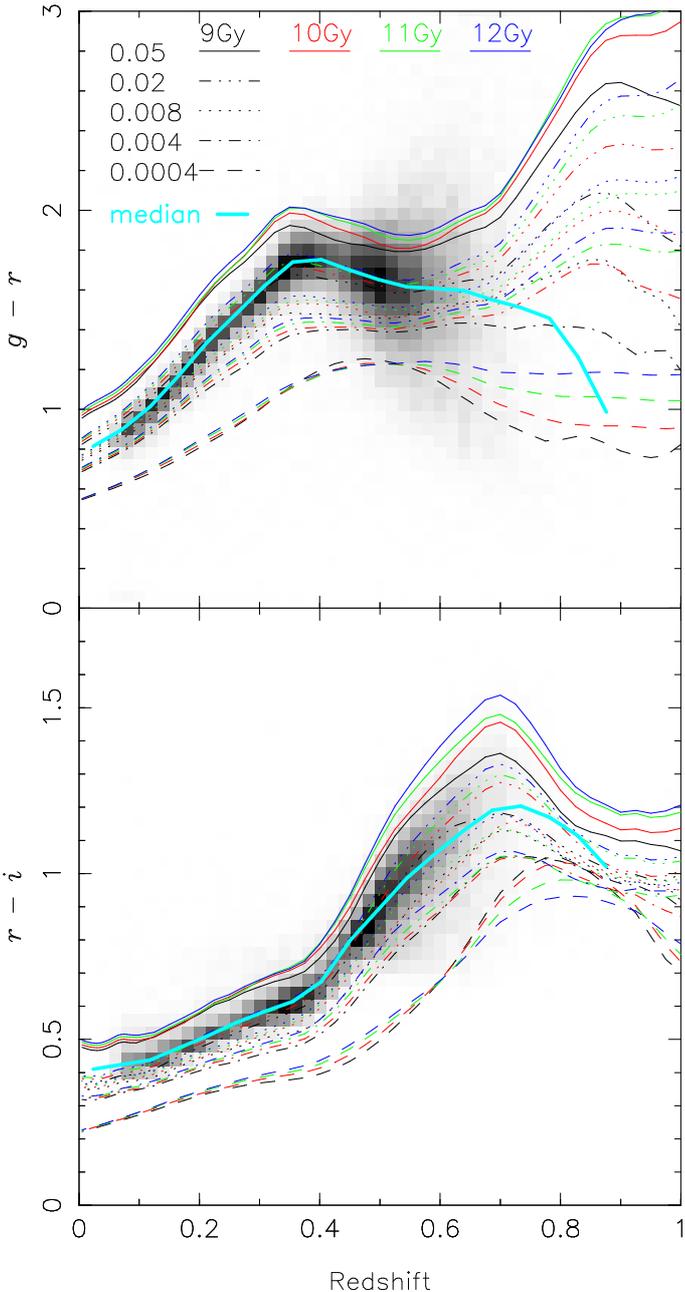}}
  \caption{The SDSS $g-r$ and $r-i$ colours versus redshift for the LRG sample used in this work. The solid cyan line indicates the median. The other lines show a range of \citet{Bruzual03} SSP models, for various formation times (different colours) and metallicities (different line-styles) as indicated in the top-left of the figure.}
  \label{plot_colev}
\end{figure}


\subsubsection{Luminosities \label{sec_data_lum}}
\begin{figure}[t!]
  \resizebox{\hsize}{!}{\includegraphics[angle=-90]{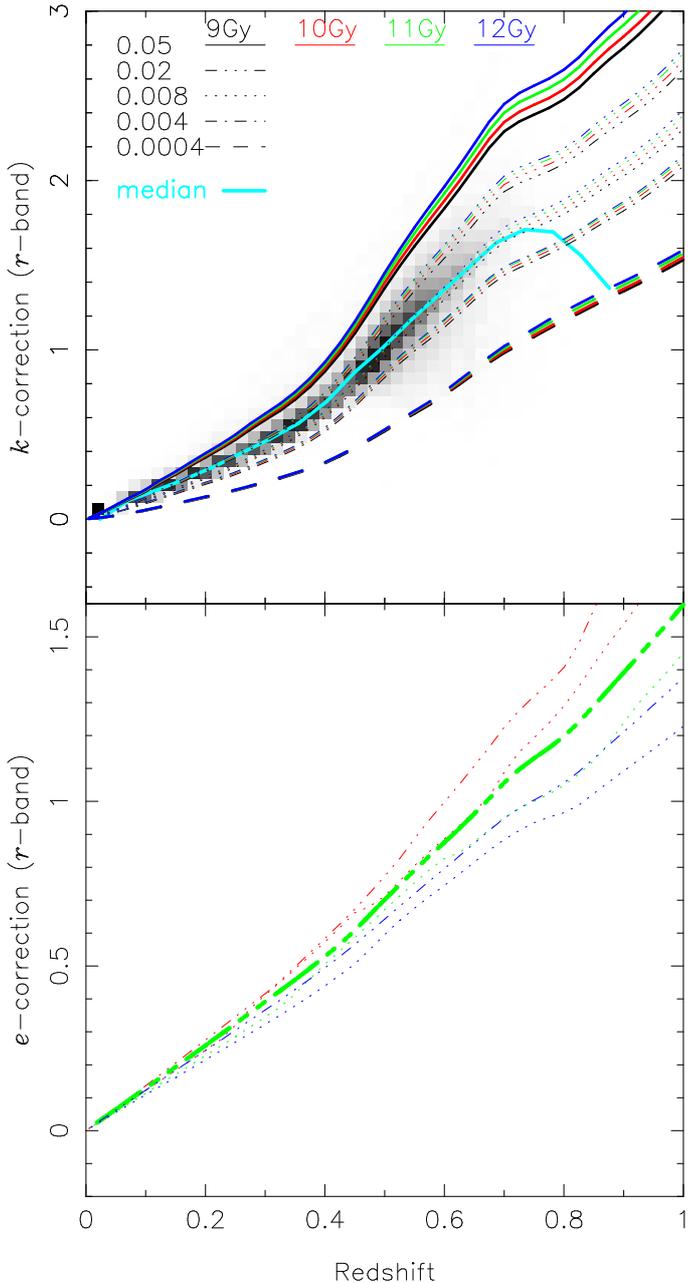}}
  \caption{({\it top panel:}) The $k$-correction as a function of redshift. The gray scale shows the $k$-corrections from the {\tt KCORRECT} code, the solid cyan line indicates the median and the other coloured lines show the $k$-correction as predicted by a range of \citet{Bruzual03} SSP models, for various formation times (different colours) and metallicities (different line-styles) as indicated in the top-left of the figure. ({\it bottom panel:}) The luminosity evolution correction as a function of redshift. For clarity, we only show a few of the \citet{Bruzual03} model predictions. The thick green dot-dot-dashed line shows the correction we have used in this work, which is based on the $Z=0.02$ instantaneous-burst model that formed 11Gyrs ago.}
  \label{plot_kecorr}
\end{figure}

\begin{figure}[t!]
  \resizebox{\hsize}{!}{\includegraphics[angle=-90]{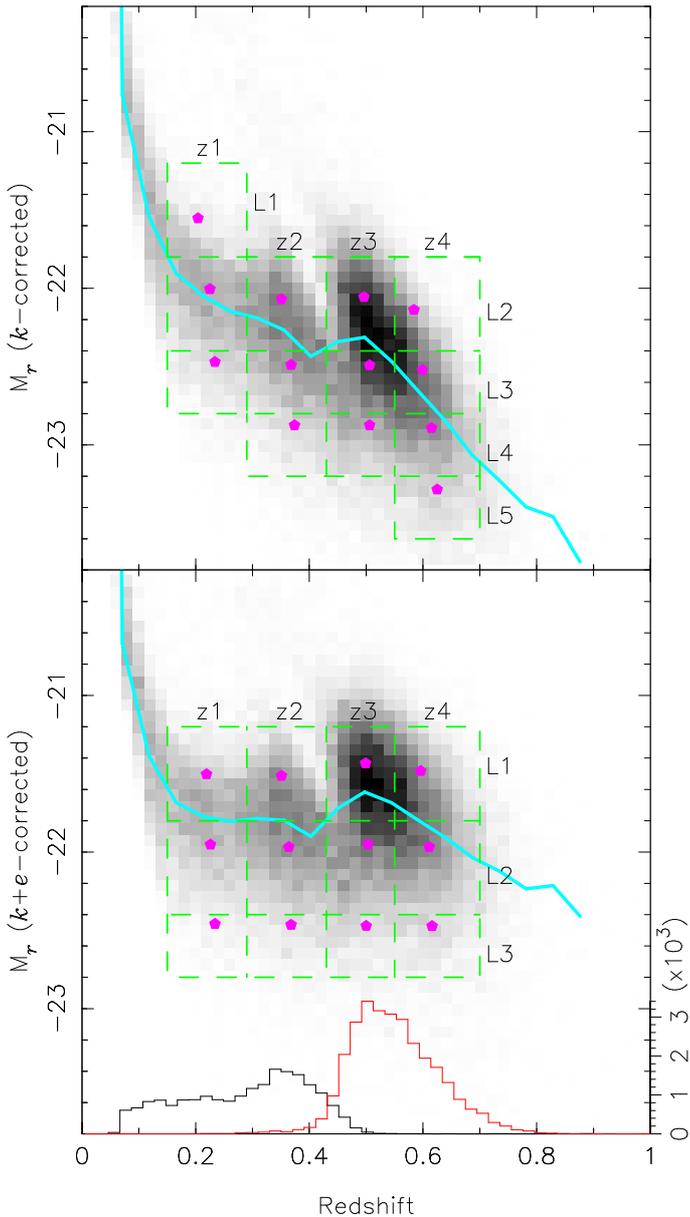}}
  \caption{The distribution of absolute magnitudes and redshifts of our LRG lenses after the $k$-correction ({\it top}) and after the $k$-correction and luminosity evolution correction ({\it bottom}). The solid cyan line indicates the median. The green dashed boxes show the redshift and luminosity cuts on our lens sample; the magenta pentagons indicate the mean redshift and luminosity of those bins. The total density of LRGs as a function of redshift is shown by the black and red histograms on the horizontal axis for the LOWZ and CMASS galaxies, respectively. }
  \label{plot_magdist}
\end{figure}

In order to study how the average halo mass of LRGs evolves, we want to compare LRGs at low redshifts to their predecessors at higher redshifts. Hence we need to obtain the luminosities of our LRGs, corrected for the redshift of their spectra through the passbands (i.e. the $k$-correction). We compute the $k$-correction using the {\tt KCORRECT} v4\_2 code \citep{Blanton03,BlantonR07}, where we use the $u$, $g$, $r$, $i$ and $z$ model magnitudes and the spectroscopic redshift as input. Furthermore, we correct for the intrinsic evolution of the luminosities (the $e$-correction), accounting for the difference between the observer-frame absolute magnitude of a galaxy with and without an evolving spectrum. \\
\indent The luminosities of LRGs are thought to evolve passively, which can be modeled using a stellar population synthesis code. We make use of one of the publicly available codes, {\tt GALAXEV} \citep{Bruzual03}, in the default configuration, i.e. adopting a \citet{Chabrier03} IMF and using the Padova 1994 tracks for the stellar evolution. We compute a range of instantaneous-burst models, where we vary the formation time and the metallicity. In Figure \ref{plot_colev}, we show the evolution of the $g-r$ and $r-i$ colours of these models, together with the observed colours of the LRGs. The set of models that describe the data best are those that assume a metallicity of $Z=0.02$ ($Z_\odot$). However, at $z<0.4$, the observed $g-r$ colours are a bit too red, and at $0.4<z<0.7$ the $r-i$ colours are somewhat too red. \citet{Maraston09} improved the modelling by including a very low metallicity component to the model that consisted of 3\% in mass, and by using the \citet{Pickles98} empirical spectral library instead of the theoretical one. However, below a redshift of $\sim$0.5 the evolution correction is fairly insensitive to the details of the modelling (see Figure \ref{plot_kecorr}), while at higher redshifts it is not clear whether the changes from \citet{Maraston09} improve the match due to the low number of objects at this redshift range used in that work. As we will discuss below, our results do not critically depend on the choice of the model, hence we do not deem it necessary to include the improvements from \citet{Maraston09}.  \\
\indent In Figure \ref{plot_kecorr} we show the $k$-correction that these {\tt GALAXEV}  models predict, together with the $k$-correction for the LRGs that have been computed using {\tt KCORRECT}. We find that at $z<0.4$, the $Z=0.02$ tracks agree well, but at higher redshifts the $k$-correction values from {\tt KCORRECT} are somewhat lower than the {\tt GALAXEV} model. In fact, the agreement at $0.4<z<0.7$ with the $Z=0.008$ metallicity models is remarkable, but the validity of these models for our LRGs at low redshift is excluded based on the colour evolution in Figure \ref{plot_colev}. However, at $z>0.4$ the LRGs show an increasing scatter in their colours and become more compatible with the $Z=0.008$ models. \\
\indent We show the luminosity evolution of some of the {\tt GALAXEV} models in Figure \ref{plot_kecorr}. We only show the models with $Z=0.02$ and $Z=0.008$, as the models with different metallicities are excluded based on their colour evolution and $k$-corrections. Also, we only show models with a formation time of 10, 11 and 12 Gyrs, as most previous works on the luminosity evolution of LRGs have adopted a formation time in this range \citep[e.g][]{Wake06,Maraston09,Banerji10,Carson10,Liu12}. For our nominal luminosity evolution correction, we adopt the $Z=0.02$ model that formed 11 Gyrs ago (at redshift 0). As none of the models exactly captures the trends in Figure \ref{plot_colev} and \ref{plot_kecorr}, the evolution correction we use may have a small bias. However, we have also tried different evolution corrections, with corresponding models that broadly cover the observed colour evolution and $k$-correction values. We detail on this test in Section \ref{sec_sens}; the main result is that our results do not change significantly. This suggests that the systematic bias in the luminosities caused by an incorrect evolution correction is likely insignificant for this work. \\
\indent LRGs have formed over a certain range of time and with some range of metallicities. Hence their actual luminosity evolution corrections may have some scatter compared to our nominal correction, as we found that the luminosity evolution correction is increasingly sensitive with redshift to these parameters. If this scatter is random with respect to our nominal correction, this causes an Eddington bias, as lenses are preferentially scattered to where there are fewer of them. In Appendix \ref{sec_ap_int}, we estimate the impact this may have on our masses estimates. We find that it is significantly smaller than our statistical errors and can be safely ignored. \\
\indent In Figure \ref{plot_magdist}, we show the distribution of absolute magnitudes after including the $k$-correction and the ($k+e$)-correction. In the range $0.15<z<0.65$ the distribution of $k+e$ corrected absolute magnitudes is fairly flat. At redshifts $z<0.15$ we have a tail of fainter objects in our catalogues, which are likely different types of galaxies. Therefore, we exclude them from this analysis. At higher redshifts, we start loosing fainter objects due to incompleteness. Since we determine both the mean luminosity and the mean halo mass for a given lens sample, this should not bias the overall mass-to-luminosity relation. \\ 


\subsection{Lensing measurement \label{sec_data_lens}}
The shapes of the background galaxies are measured on images from the RCS2. Details on the data reduction and the shape measurement process can be found in \citet{VanUitert11}, and some important improvements of our lensing analysis were discussed in \citet{Cacciato14}. It suffices to say that we measure the shapes of the galaxies with the KSB method \citep{Kaiser95,LuppinoK97,Hoekstra98}, using the implementation described by \citet{Hoekstra98,Hoekstra00}. This method was tested on a range of simulations as part of the Shear Testing Programme (STEP) 1 and 2 (the `HH' method in Heymans et al. 2006 and Massey et al. 2007 respectively) where it was found to have a multiplicative bias of a few per cent at most and a negligible additive bias. Recently, \citet{Hoekstra15} found that these results were driven by the overly simplistic nature of the STEP simulations; for more realistic simulations, KSB suffers from noise bias \citep{Kacprzak12,Melchior12,Refregier12} as any other shape measurement method that is currently in use. We calibrate our KSB implementation on realistic image simulations generated with GalSim\footnote{https://github.com/GalSim-developers/GalSim} \citep{Rowe14} that are set up to closely match RCS2 observations, i.e. with an intrinsic ellipticity distribution that is matched to the observations, a realistic range of S\'ersic profile indices for the simulated galaxies, and up to a magnitude limit that is matched to the RCS2. We determine the multiplicative bias as a function of seeing:
\begin{equation}
m_{\rm corr}=-0.065\times({\rm FWHM}-0.7)-0.123,
\end{equation}
with FWHM the size of stars in an image. We use this to correct the shear measured in each RCS2 image. We do not need to correct for residual additive bias as that generally averages out in galaxy-galaxy lensing due to symmetry in lens-source pair orientations. The (multiplicative) noise bias correction increases the average lensing signal by \mbox{10-15\%}. Note that the correction is not very sensitive to the adopted width of the intrinsic ellipticity distribution, but that it is critically important to simulate galaxies up to $\sim$1.5 magnitudes deeper than the nominal magnitude limit of the survey \citep[see][]{Hoekstra15}. \\
\indent The lensing signal is extracted by azimuthally averaging the tangential projections of the ellipticities of the source galaxies in concentric radial bins, i.e. by measuring the tangential shear as a function of projected separation:
\begin{equation}
  \langle\gamma_t\rangle(R) = \frac{\Delta\Sigma(R)}{\Sigma_{\mathrm{crit}}},
\end{equation}
where $\Delta\Sigma(R)=\bar{\Sigma}(<R)-\bar{\Sigma}(R)$ is the difference between the mean projected surface density inside radius $R$ and the projected surface density at $R$, and $\Sigma_{\mathrm{crit}}$ is the critical surface density:
\begin{equation}
  \Sigma_{\mathrm{crit}}=\frac{c^2}{4\pi G}\frac{D_{\rm s}}{D_{\rm l}D_{\rm ls}},
\end{equation}
with $D_{\rm l}$, $D_{\rm s}$, and $D_{\rm ls}$ the angular diameter distance to the lens, the source, and between the lens and the source, respectively. All galaxies with an apparent magnitude of $22<r'<24$ and a well-defined shape measurement are selected as sources. \\
\indent We measure the lensing signal from the BOSS lenses in each \mbox{1$\times$1 deg$^2$} RCS2 pointing, including the sources from the neighbouring pointings (if present). We bootstrap over these patches to obtain the covariance matrix, which accounts for intrinsic shape noise, measurement noise, as well as the contribution from large-scale structures. The off-diagonal elements are consistent with zero on the radial range of interest ($<2h_{70}^{-1}$Mpc), hence we only use the inverse of the diagonal as the errors on the measurement when fitting the models to the data.\\
\indent As these patches overlap, the contribution from large-scale structures might be somewhat underestimated at large scales; therefore, as a test, we also perform the lensing measurements on \mbox{2$\times$2 deg$^2$} non-overlapping patches and use that in the bootstrap resampling. We find that for both methods, the signal and the covariance matrix barely change in the radial range that we use in this work. Since the total area decreases if we limit our analysis to \mbox{2$\times$2 deg$^2$} non-overlapping patches only, and since it makes no difference to the signal and its error, we decide to use all \mbox{1$\times$1 deg$^2$} RCS2 pointings plus neighbours as basis for the bootstrapping.  \\
\indent In order to compute $\Sigma_{\mathrm{crit}}$ we need the distances to the lenses and sources. We compute $D_{\rm l}$ for each lens separately using its spectroscopic redshift from SDSS. The lensing efficiencies $\langle D_{\rm ls}/D_{\rm s} \rangle$ are determined by averaging over the source redshift distribution, which is obtained by applying the same $r'$-band selection to the photometric redshift catalogues of the COSMOS field from \citet{Ilbert13}. The procedure is described in more detail in Appendix C of \citet{Cacciato14}. Note that we previously used the photometric redshift catalogues from \citet{Ilbert09} as the former was not yet publicly available. Comparing the average lensing efficiencies from the two catalogues, we find them to agree for low lens redshifts, but to be increasingly different at higher redshifts (up to 15\% at z$_{\rm l}$=0.7). At increasingly high lens redshift the lensing efficiencies are more sensitive to the form of the adopted source redshift distribution, which is somewhat different for the two catalogues. We discuss the robustness of the derived lensing efficiencies in more detail in Appendix B. \\
\indent A fraction of the sources is physically associated to the lens galaxies, representing an overdensity of sources galaxies that are not lensed. We cannot remove them since we lack redshifts for our sources. Such a contamination in the source catalogue dilutes the lensing signal. This can be corrected for by measuring the excess source number density relative to the background as a function of projected separation, and boosting the lensing signal with this factor \citep[e.g.][]{Mandelbaum06,VanUitert11}. We follow the same procedure. \\ 
\indent This boost correction itself is biased low as the galaxies associated to the lens, and the lens itself, block light from the background sky, suppressing the source number density. The effect is described in \citet{Simet14}. As discussed in that work, a correction for this bias is obtained by multiplying the boost correction with a factor $1/(1-f_{\rm obsc})$, where $f_{\rm obsc}$ is the fraction of the sky that is obscured by the foreground galaxies. We compute this by using the ISOAREA\_IMAGE keyword in {\tt SExtractor}, which stores how many pixels a galaxy spans on the sky. $f_{\rm obsc}$ is taken to be the sum of these values of all galaxies whose centroids fall inside a radial bin, divided by the total number of pixels in that bin (accounting for the effect of the survey masks and geometry). Before doing the correction, we subtract the average sky-background of $f_{\rm obsc}$ from the one observed around the lenses, as we are only interested in the additional obscuration. We find that the correction is at most 5\%, in the radial bins closest to the most luminous and low-redshift lenses. The correction decreases at larger radii, as well as for fainter, higher redshift LRGs, as expected. \\
\indent The robustness of the lensing signal has been addressed in Appendix B of \citet{Cacciato14}. There, we show that the cross shear signal of our lens sample is consistent with zero. Also, the random shear signal, which is used to correct for the effect of residual systematics in the shape measurement catalogues, is smaller than the real signal for all the projected separations we use in this work. However, an overall multiplicative bias could still be present, either through an incorrect determination of the noise bias correction, or through the use of incorrect lensing efficiencies. In Appendix B we perform an internal consistency check of our measurement pipeline, which provides strong evidence that such a bias is unlikely to be significant. 


\section{Halo model \label{sec_hm}}
In this section we describe the model that we employ to provide a physical interpretation of our measurements. The halo model provides a useful framework to describe the stacked weak lensing signal around galaxies \citep[see e.g.][]{Mandelbaum06,Cacciato09,Cacciato14,Miyatake13}. It is based on a statistical description of dark matter properties, such as their average density profile, their abundance, and their large scale bias, complemented with a statistical description of the way galaxies of a given luminosity populate dark matter haloes of different masses  (also known as halo occupation statistics). In its fundamental assumptions, the model is similar to the one presented in \citet{Seljak00}, \citet{CoorayS02} and \citet{Cacciato09}. \\
\indent Galaxy-galaxy lensing probes the average matter distribution projected along the line-of-sight at a given projected physical separation, $R$, for a set of lenses. The quantity of interest is the excess surface mass density profile, \mbox{$\Delta\Sigma(R)$}, which is determined from the projected surface mass density, $\Sigma(R)$. Since we measure the average signal of many lenses, the projected matter density can be expressed in terms of the galaxy-dark matter cross-correlation, $\xi_{\rm gm}(r)$:
\begin{equation}
\Sigma(\theta) = \bar{\rho}_\rmm \int_{0}^{\omega_{\rm s}}
\left[1+\xi_{\rm gm}(r)\right] \rmd \omega,
\label{Sigma_approx}
\end{equation}
where the integral is along the line of sight, $\omega$ is the comoving distance from the observer, $\omega_{\rm s}$ the comoving distance to the source and $\bar{\rho}_\rmm$ is the mean matter density at the redshift of the lens. The three-dimensional comoving distance $r$ is related to $\omega$ through $r^2 = \omega_{\rm l}^2  + \omega^2 - 2 \omega_{\rm l}\omega \cos \theta$, with $\omega_{\rm l}$ the comoving distance to the lens and $\theta=R/D_{\rm l}$ the angular separation between lens and source (see Figure 1 in \citet{Cacciato09}). Note that the galaxy-dark matter cross-correlation is evaluated at the average redshift of the lens galaxies. \\
\indent Under the assumption that each galaxy resides in a dark matter halo, $\Delta\Sigma$ can be computed using a statistical description of how galaxies are distributed over dark matter haloes of different masses \citep[see e.g.][]{VanDenBosch13}. Specifically, it is fairly straightforward to obtain the two-point correlation function, $\xi_{\rm gm}(r,z)$, by Fourier transforming the galaxy-dark matter power-spectrum, $P_{\rm gm}(k,z) $, i.e.
\begin{eqnarray}
\xi_{\rm gm}(r,z) = {1 \over 2 \pi^2} \int_0^{\infty} P_{\rm gm}(k,z)
{\sin kr \over kr} \, k^2 \, \rmd k\,,
\label{xiFTfromPK}
\end{eqnarray}
with $k$ the wavenumber. The quantity $P_{\rm gm}(k,z)$ can be expressed as a sum of a term that describes the small scales (one-halo, 1h) and one that describes the large scales (two-halo, 2h), each of which can be further subdivided based upon the type of the galaxies (central or satellite) that contribute to the power spectrum. This reads
\begin{equation}
P_{\rm gm}(k) = P^{\rm 1h}_{\rm cm}(k) + P^{\rm 1h}_{\rm sm}(k) +
P^{\rm 2h}_{\rm cm}(k) + P^{\rm 2h}_{\rm sm}(k)\,.
\label{eq_powerspec}
\end{equation}
The terms in Equation \ref{eq_powerspec} can be written in compact form as
\begin{equation}
\label{P1h}
P^{\rm 1h}_{\rm xy}(k,z) = \int \calH_\rmx(k,M,z) \, \calH_\rmy(k,M,z)
\, n_{\rm h}(M,z) \, \rmd M,
\end{equation}
\begin{eqnarray}
\label{P2h}
\lefteqn{P^{\rm 2h}_{\rmx\rmy}(k,z) = \int \rmd M_1 \,
\calH_\rmx(k,M_1,z) \, n_{\rm h}(M_1,z) } \nonumber \\
& & \int \rmd M_2 \, \calH_\rmy(k,M_2,z) \, n_{\rm h}(M_2,z) \,
Q(k|M_1,M_2,z)\,,
\end{eqnarray}
where `x' and `y' are either `c' (for central), `s' (for satellite), or `m' (for matter), \mbox{$Q(k|M_1,M_2,z)= b_h(M_1,z)b_h(M_2,z)P^{\rm lin}_{\rm m}(k,z)$} describes the power spectrum of haloes of mass $M_1$ and $M_2$,  and it contains the large scale bias of haloes $b_h(M)$ from \cite{Tinker10} (but see \citet{VanDenBosch13} for a more sophisticated modelling of this term).  $n_{\rm h}(M,z)$ is the halo mass function of \cite{Tinker10}. Furthermore, we have defined
\begin{equation}
\calH_\rmm(k,M,z) = {M \over \bar{\rho}_{\rmm}} \, \tilde{u}_\rmh(k|M,z)\,,
\label{calHm}
\end{equation}
and
\begin{equation}
\label{calHc}
\calH_\rmx(k,M,z) =
{\langle N_\rmx|M \rangle \over \bar{n}_{\rmx}(z)}\,
{\tilde u}_\rmx(k|M)
\end{equation}
where
\begin{equation}
{\tilde u}_\rmc(k|M) = 1-p_{\rm off}+p_{\rm off}\times{\rm exp}[-0.5k^2(r_s\{M\}{\cal R}_{\rm off})^2] \, ,
\label{eq_poff}
\end{equation}
and
\begin{equation}
\tilde{u}_\rms(k|M,z) = \tilde{u}_\rmh(k|M,z)\,.
\end{equation}
$p_{\rm off}$ is the parameter that describes the probability that the `central' galaxy does not reside at the centre of the dark matter halo, whereas ${\cal R}_{\rm off}$ quantifies the amount of off-centering in terms of the halo scale radius, $r_s(M)$ \citep[see e.g.][]{Skibba11,More14}. In our fiducial model, we set $p_{\rm off} = {\cal R}_{\rm off}= 0$, but we explore the impact of this assumption in Section \ref{sec_res}. The functions $\langle N_{\rm c}|M \rangle$ and $\langle N_{\rm s}|M \rangle$ represent the average number of central and satellite galaxies in a halo of mass $M \equiv 4 \pi(200 {\bar \rho_\rmm})R_{200}^3/3$, defined as:
\begin{equation}
\langle N_{\rm c}|M \rangle=  \frac{1}{\sqrt{2 \pi} {\rm ln}(10) M\sigma_{\rm \log M} }
\exp\left( -\frac{(\log M - \log M_{\rm mean})^2}{2\sigma_{\rm \log M}^2}\right) 
\label{eq_nofm}
\end{equation}
\begin{equation}
\langle N_{\rm s}|M \rangle = (M/M_1)\, f_{\rm trans}(M) \,
\end{equation}
where
\begin{equation}
 f_{\rm trans}(M) =  0.5 \times \left[ 1+{\rm erf} \left( \frac{\log M -\log{M_{\rm cut}})}{\sigma_{\rm trans}} \right) \right].
\end{equation}
We use a flat, non-informative prior for $\sigma_{\rm \log M}$  and $M_{\rm mean}$, set $M_{\rm cut} = \langle M_{\rm eff} \rangle$ (see Equation \ref{eq:haloMasses}) and $\sigma_{\rm trans}=0.25$. Since LRGs are thought to be predominantly central galaxies \citep[see e.g.][]{Wake08,Zheng09,Parejko13}, we set $\langle N_\rms|M \rangle$ to zero and only fit for $M_{\rm mean}$ and $\sigma_{\rm M}$ in our nominal runs. We test the impact of this assumption on the derived quantities in the result sections by additionally fitting for $M_1$. We have tested that the result is fairly insensitive to the details of the modelling of $f_{\rm trans}(M)$. \\
\indent $\bar{n}_{\rmg}(z)$ is the number density of galaxies at redshift $z$:
\begin{eqnarray}
\bar{n}_{\rmg}(z) &=&
\int \langle N_{\rm g}|M \rangle n_{\rm h}(M,z) {\rm d} M
\nonumber \\
& \approx &
\int \langle N_{\rm c}|M \rangle n_{\rm h}(M,z) {\rm d} M
\, .
\end{eqnarray}
The last equality is exact in the case of LRGs being only central galaxies. $\tilde{u}_\rmh(k|M,z)$ is the Fourier transform of the normalized density distribution of dark matter within a halo of mass $M$, for which we assume a Navarro-Frenk-White (NFW) profile \citep{Navarro96} and a mass-concentration relation from \citet{Duffy08}:
\begin{equation}
c_{\rm m} = A\left(\frac{M}{M_{\rm pivot}}\right)^{B} (1+z)^{C}\, ,
\label{eq_mc}
\end{equation}
with $A = f_{\rm conc} \times 10.14$, $B = -0.081$, $C=-1.01$, and $M_{\rm pivot} = 2\times 10^{12} h^{-1} M_\odot$. Note that $f_{\rm conc}$ is a free parameter that allows the normalisation of this relation to vary. Specifically, we apply a non-informative flat prior on this parameter. \\
\indent The average halo mass in a given luminosity bin, which is what we shall refer to as `effective' halo mass in what follows, can then be computing taking into account the weight of the halo mass function:
\begin{equation}
\label{eq:haloMasses}
\langle M_{\rm eff} \rangle =\frac{\int \langle N_{\rm c}|M' \rangle
n_h(M',z_{\rm lens}) M'\mathrm{d}M'}{\int \langle N_{\rm c}|M' \rangle
n_h(M',z_{\rm lens})\mathrm{d}M'}
\end{equation}
where $z_{\rm lens}$ is the mean redshift of the lens galaxies in a given luminosity bin. The distinction between the mass associated with the mean of the log-normal distribution and the halo mass inferred accounting for the mass function is of relevance because LRGs populate fairly massive haloes for which the mass function is steep (see e.g. Figure 7 in \citet{Leauthaud15}).  \\
\indent At small scales one expects the baryonic mass of LRGs to contribute to the lensing signal. The smallest scale used in this study is 50 $h^{-1}_{70}$kpc, which is much larger than the typical extent of the baryonic content of a galaxy. Therefore, it is adequate to model the lensing signal of the LRGs itself as a point source of mass $M_{\rm g} \approx M_*$. This reads
\begin{equation}
\Delta \Sigma^{\rm 1h, g}(R) \approx \frac{\langle M_* \rangle_{L_-}^{L_+} }{\pi \, R^2} \, .
\end{equation}
We use the value of the average stellar masses, $\langle M_{*}\rangle$, for the galaxies in the luminosity bins under investigation here. The stellar masses are obtained by matching our lens catalogue to the MPA-JHU stellar mass catalogue\footnote{http://www.mpa-garching.mpg.de/SDSS/DR7/}. As the MPA-JHU catalogue is based on the MAIN sample from SDSS, we only have matches at low redshift. However, the point mass has a small impact on our fit results; hence we do not expect that a potential evolution of the average stellar mass-to-light ratio for LRGs can be so strong that it could significantly affect our results. \\
\indent To summarize, our fiducial model for the lensing signal is the sum of three terms: one describing the lensing due to the baryonic mass; the second is responsible for the small (sub-Mpc) scale signal mostly due to the dark matter density profile of haloes hosting central LRGs; and the last describes the large (a few Mpc) scale signal due to the clustering of dark matter haloes around LRGs. This reads:
\begin{eqnarray}
\label{eq:esdterms}
\Delta \Sigma(R) &=& \Delta \Sigma^{\rm 1h,g}(R) +\Delta \Sigma^{\rm 1h}_{\rm cm}(R) + \Delta \Sigma^{\rm 2h}_{\rm cm}(R) \, .
\label{eq:esdterms}
\end{eqnarray}
We simultaneously fit the halo model to the three luminosity bins and do this for each redshift slice separately. We have five free parameters in each fit: the three mean masses of the luminosity bins, the scatter and the normalisation of the mass-concentration relation. The fit is performed using a Markov Chain Monte Carlo Method (MCMC). Details of its implementation can be found in Appendix \ref{ap_mcmc}.  \\
\indent We fit the model to the measurements on scales between 0.05 and 2 $h^{-1}_{70}$ Mpc. At these scales, both the measured lensing signal and the halo model predictions are fairly robust. At larger scales, the lensing signal becomes smaller and more susceptible to residual systematics. At scales smaller than 0.05 $h^{-1}_{70}$Mpc, lens light may bias the shape measurements. For the halo model, the overlap between the 1-halo and 2-halo term is notoriously hard to model because of, amongst others, halo exclusion and non-linear biasing. This mainly impacts the few Mpc regime. Most of the information about the halo masses and concentrations is contained in the lensing signal within the virial radius, so we do not loose much statistical precision by limiting ourselves to these scales.

 
\section{Luminosity-to-halo mass relation \label{sec_res}}

To study how the luminosity-to-halo mass relation of LRGs evolves with redshift, we divide our sample in bins of ($k+e$)-corrected absolute magnitude and redshift as detailed in Table \ref{tab_fit2} and Figure \ref{plot_magdist}. For $z<0.43$, we only select lenses from the LOWZ sample, at higher redshifts we exclusively select CMASS galaxies. The average log stellar masses for the consecutive luminosity bins are 11.2, 11.5 and 11.7 [$\log(h^{-2}_{70}M_\odot)$]. For each bin we measure the average lensing signal, which is shown in Figure \ref{plot_gg2}, together with the best-fit halo models and the model uncertainties (computed as detailed in Appendix \ref{ap_mcmc}). We find $\chi^2_{\rm red}$ values of 1.8, 1.6, 1.2 and 1.0, going from the lowest to the highest redshift slice. Hence the fits of the CMASS samples are good, but for the LOWZ samples the $\chi^2_{\rm red}$ values are somewhat large, suggesting that either our error bars are underestimated, or that the model that we fit to the data is overly simplistic. \\
\indent The errors on the lensing measurements account for intrinsic shape noise, measurement noise and the impact of large-scale structures. We have, however, ignored some small sources of error, as their amplitude is much smaller than the statistical errors on the lensing signal: the error on the boost correction, which is typically a few percent at small scales; the error on the obscuration correction, which is even smaller; the error on determining the lensing efficiency and the error on the multiplicative bias calibration, whose magnitudes are unknown but are likely of the order of a few percent. Combined, they might increase the errors by as much as $\sim$10\%, although the exact number is difficult to estimate reliably. If we increase our error bars by this amount, we would get $\chi^2_{\rm red}$ values of 1.5, 1.4, 1.0 and 0.8, respectively. The fact, however, that we find reasonable $\chi^2_{\rm red}$ values for the CMASS sample, but not for the LOWZ sample, suggests that a systematic underestimate of our errors is unlikely to be the dominant cause. \\
\indent Even though a visual inspection of the covariance matrix lead us to believe that it is diagonal on scales \mbox{$<$2$h_{70}^{-1}$ Mpc}, there could be low-level off-diagonal terms present that, if included, would lower the $\chi_{\rm red}^2$ values. This potentially affects the LOWZ results more, as the measurements have a higher signal-to-noise ratio and the covariance matrix is less noisy. To test this, we recompute the $\chi^2_{\rm red}$ values using the full covariance matrix which we obtained from bootstrapping (see Section \ref{sec_data_lens}) for the best-fit models. Note that we only include the covariance between radial bins of a lens sample, but not the covariance between the radial bins of the different luminosity samples. If present, they would lower the $\chi_{\rm red}^2$ values even more. We find that $\chi^2_{\rm red}$ of the first redshift slice reduces to 1.6, while it does not change for the other three redshift slices. Hence the effect is small and does not fully explain the high $\chi^2_{\rm red}$ values. \\
\indent Figure \ref{plot_gg2} shows that the signal-to-noise ratio of the lensing measurements of the LOWZ samples is very high and would allow for a more sophisticated modelling. When we include a satellite term or a miscentering term in the halo model, however, the $\chi^2_{\rm red}$ values do not improve, as the lensing signal alone cannot constrain the miscentering distribution parameters very well, and the expected number of satellites is low. Allowing for even more freedom in the fit might lead to overfitting of the CMASS results. Using different halo models for the different samples, or splitting the LOWZ sample up in more luminosity bins, reduces the homogeneity of the analysis, which is one of the key advantages of our work. Hence we choose to stick to the settings described above. In Section \ref{sec_sens} we perform a sensitivity analysis and find that our results do not critically depend on various choices in the analysis, suggesting that the quantities we derive from the fits are robust. \\
\begin{table}
\renewcommand{\tabcolsep}{0.09cm}
  \caption{Properties of the lens bins (after ($k+e$)-correction)}   
  \centering
  \begin{tabular}{c c c c c c c c c c} 
  \hline  \hline
  & & & & & & & &\\
 & M$_r$ & N$_{\rm lens}$ & $\langle z \rangle$ & $\langle L_r \rangle$ & $M_{\rm eff}$ & $f_{\rm conc}$ & $\chi^2_{\rm red}$\\ 
 & (1) & (2) & (3) & (4) & (5) & (6) & (7)\\
  & & & & & & & \\
  \hline
  & & & & & & & \\
\multicolumn{3}{c}{0.15$<$$z$$<$0.29 (LOWZ)} \\
L1z1 & [-21.8,-21.2] &  2969 & 0.219 & 0.65 & $3.65_{-0.50}^{+0.52}$ & & \\  
L2z1 & [-22.4,-21.8] &  2606 & 0.226 & 0.99 & $5.60_{-0.69}^{+0.73}$ & $0.68^{+0.08}_{-0.06}$ & 1.8 \\ 
L3z1 & [-22.8,-22.4] &   300 & 0.234 & 1.58 & $14.9_{-2.2}^{+2.4}$  & & \vspace{2mm} \\  
\multicolumn{3}{c}{0.29$<$$z$$<$0.43 (LOWZ)} \\
L1z2 & [-21.8,-21.2] &  3771 & 0.351 & 0.66 & $2.85_{-0.45}^{+0.49}$ & & \\  
L2z2 & [-22.4,-21.8] &  4502 & 0.364 & 1.00 & $5.15_{-0.64}^{+0.69}$ &$0.85^{+0.12}_{-0.11}$ & 1.6 \\
L3z2 & [-22.8,-22.4] &   721 & 0.368 & 1.58 & $9.86_{-1.50}^{+1.66}$ & & \vspace{2mm} \\  
\multicolumn{3}{c}{0.43$<$$z$$<$0.55 (CMASS)} \\
L1z3 & [-21.8,-21.2] &  8530 & 0.499 & 0.61 & $2.03_{-0.39}^{+0.43}$ & & \\  
L2z3 & [-22.4,-21.8] &  4213 & 0.503 & 0.99 & $4.67_{-0.77}^{+0.88}$ &$0.77^{+0.19}_{-0.15}$ & 1.2 \\ 
L3z3 & [-22.8,-22.4] &  587 & 0.500 & 1.59 & $6.52_{-1.66}^{+1.91}$ & & \vspace{2mm} \\  
\multicolumn{3}{c}{0.55$<$$z$$<$0.70 (CMASS)} \\
L1z4 & [-21.8,-21.2] &  5256 & 0.596 & 0.64 & $1.92_{-0.57}^{+0.66}$ & & \\  
L2z4 & [-22.4,-21.8] &  5161 & 0.611 & 1.00 & $4.16_{-0.90}^{+1.01}$ &$0.73^{+0.25}_{-0.20}$ & 1.0 \\ 
L3z4 & [-22.8,-22.4] &  969 & 0.616  & 1.60 & $6.48_{-1.78}^{+2.11}$ & & \\  
  & & & & & & & & \\
  \hline \hline
 & & \\
  \end{tabular}
  \tablefoot{(1) absolute magnitude range (after ($k+e$)-correction); (2) number of lenses; (3) mean redshift; (4) mean luminosity \mbox{[10$^{11} h_{70}^{-2} L_\odot$]} (after ($k+e$)-correction); (5) best-fit halo mass \mbox{[10$^{13} h_{70}^{-1} M_\odot$]}; (6) best-fit normalisation of the mass-concentration relation; (7) reduced chi-squared of the fit.}
  \label{tab_fit2}
\end{table} 
\begin{figure*}[t!]
  \resizebox{\hsize}{!}{\includegraphics{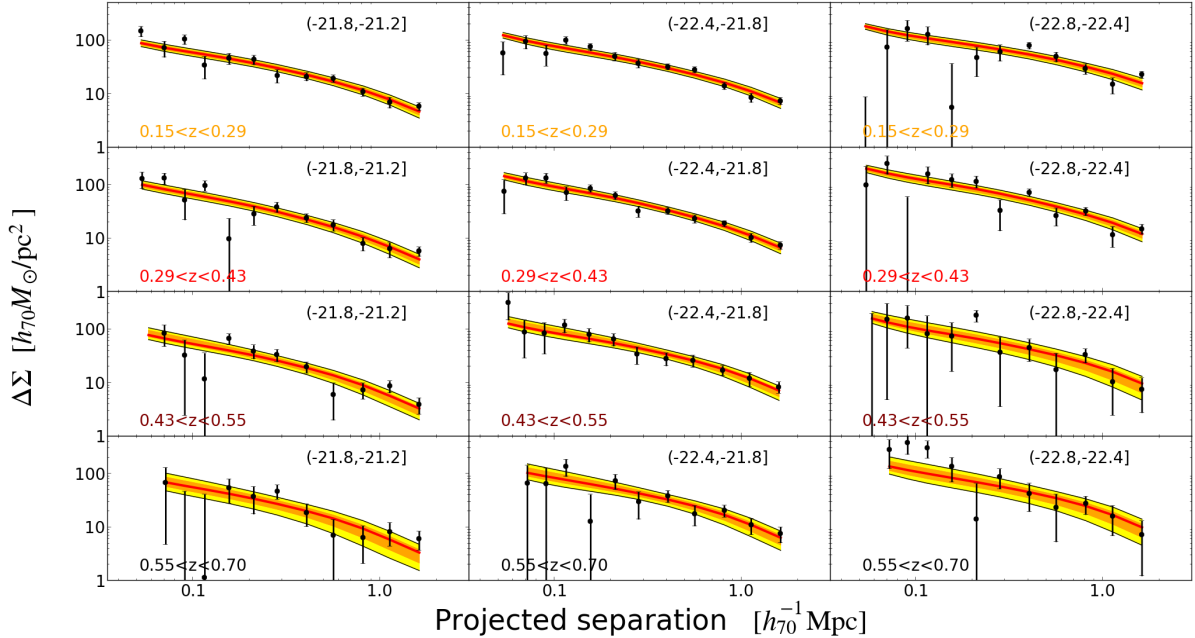}}
  \caption{The lensing signal $\Delta \Sigma$ of LOWZ (top two rows) and CMASS (bottom two rows) lenses as a function of projected separation for the three luminosity bins (after the ($k+e$)-correction is applied). The solid red lines show the best-fit halo model, the orange and yellow regions the 1 and 2$\sigma$ model uncertainty, respectively. We fit the signal on scales between 0.05 and 2 $h^{-1}_{70}$ Mpc.}
  \label{plot_gg2}
\end{figure*}

\begin{figure}
  \resizebox{\hsize}{!}{\includegraphics{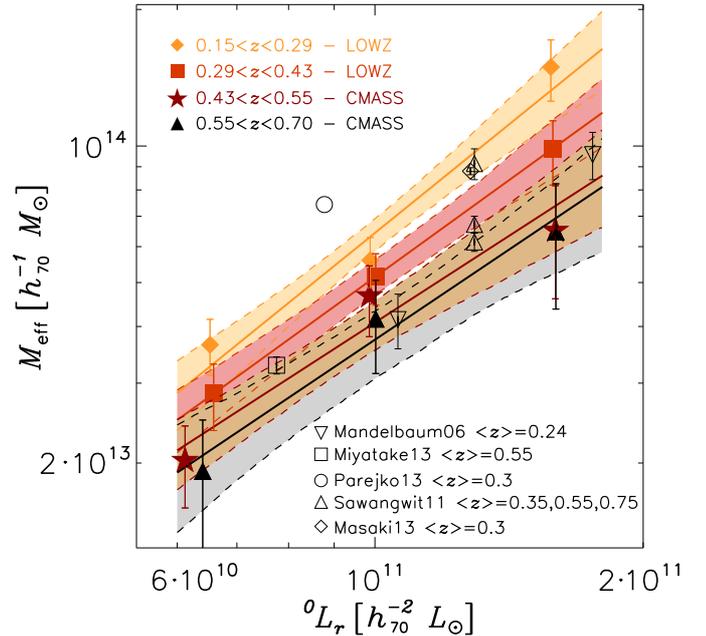}}
  \caption{The mean ($k+e$)-corrected luminosity versus the best-fit halo mass of the twelve lens samples. The different symbols indicate the different redshifts bins, as indicated in the figure. The coloured areas indicate the 68\% confidence intervals of the powerlaw fits.}
  \label{plot_lm2}
\end{figure}

\begin{figure}
  \resizebox{\hsize}{!}{\includegraphics[angle=-90]{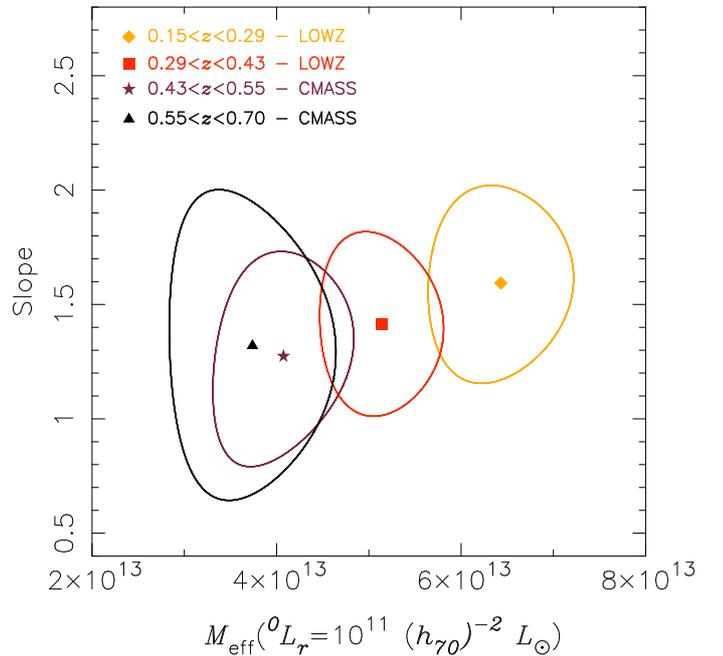}}
  \caption{68\% confidence contours for two parameters of the powerlaw fit between luminosity and halo mass.}
  \label{plot_conf2}
\end{figure}

\indent In the halo model, we fit the mean and the scatter of the log-normal distribution that describes $\langle N_\rmc|M \rangle$. The log of the mean has typical values of $\sim$14.5, $\sim$15 and $\sim$15.5 for the three luminosity bins, while the scatter ranges between 0.7 and 0.8. Neither evolves with redshift. Note that this is the scatter in the log of the halo mass and not in luminosity. The latter was fit in \citet{Cacciato14} where it was found to have a value of $\sigma_{\rm log L_c}=0.146\pm0.011$, obtained by fitting the halo model to the lensing signal of all galaxies in the DR9 that overlap with RCS2. The scatter in halo mass is much larger, because the luminosity-halo mass relation flattens at higher luminosities; a small scatter in luminosity corresponds to a large one in halo mass (see e.g. Figure 3 and the discussion in \citet{More09}). \\
\indent The quantity of interest that we can compare to other works is the `effective' halo mass, which is given in Table \ref{tab_fit2}. We plot it as a function of luminosity in Figure \ref{plot_lm2}. We find that the masses increase with luminosity and decrease with redshift. To quantify this, we parametrize the luminosity-to-halo mass relation by $M_{\rm eff}=M_{0,L}(L/L_0)^{\beta_L}$, using a pivot luminosity of $L_0=10^{11}h_{70}^{-2}L_\odot$. The best-fit powerlaw fits are shown in the same figure; the confidence contours of the fitted amplitude and slope are shown in Figure \ref{plot_conf2}. We have listed the fit parameters in Table \ref{tab_fitpar2}. \\
\indent The slope of the luminosity-to-halo mass relation does not change significantly for our different redshift samples and has a typical value of 1.4. The amplitude, however, is about $\sim$4$\sigma$ higher for our lowest redshift slice compared to the highest one. On average, the masses of LOWZ galaxies increase by $25_{-14}^{+16}$\% between redshift 0.36 and 0.22; the masses of CMASS galaxies increase by $10_{-20}^{+25}$\% from redshift 0.6 to 0.5. If we assume that CMASS galaxies evolve into LOWZ galaxies and combine the results, we find an average increase of $80_{-28}^{+39}$\% (stat. errors) in $M_{\rm eff}$ at $L_0=10^{11}h_{70}^{-2}L_\odot$ from $z$$\sim$0.6 to $z$$\sim$0.2. Fixing the slope to its average value of 1.4 and only fitting the amplitude changes this number to $77_{-27}^{+36}$\%.  \\
\indent \citet{Tojeiro12} found that at brighter absolute magnitudes, a larger fraction of CMASS galaxies are the progenitors of LOWZ galaxies. If we discard the lowest luminosity bin, we find that, for a pivot luminosity of \mbox{$L_0=1.3\times10^{11}h_{70}^{-2}L_\odot$}, the average halo mass increases with $97_{-38}^{+52}$\%. Considering only the brightest luminosity bin, the average halo mass increases with $160_{-76}^{+133}$\%. \\


\subsection{Sensitivity analysis \label{sec_sens}}
\indent To study how sensitive our results are to the adopted luminosity evolution correction, we perform the following test. We multiply our nominal correction with a factor, such that the resulting luminosity evolution curves roughly cover the range of reasonable models that are shown in the lower panel of Figure \ref{plot_kecorr}. The factors we choose are  $1-0.2\times z$ and  $1+0.2\times z$, respectively. Next, we recompute the luminosities, repeat the lensing measurements (using the same cuts) and the halo model fits, and compare the resulting best-fit halo masses. We find that the best-fit halo masses of the individual luminosity bins do not significantly shift compared to the nominal results. We fit the luminosity-to-halo mass relation and list the parameters in Table \ref {tab_fitpar2}. As is shown there, we find that the best-fit slopes are very similar. The best-fit amplitude shifts with 1$\sigma$ at most compared to our nominal results. For the $1-0.2\times z$ modification factor, the resulting increase in halo mass of LRGs from $z=0.6$ to $z=0.2$ is $85^{+43}_{-30}$\%; for the $1+0.2\times z$ modification factor, it is $50^{+30}_{-23}$\%. The halo masses and corresponding growth change somewhat because the lens selection shifts systematically, such that we analyse lens samples that to some extent are intrinsically different. Importantly, however, our results do not critically depend on the choice of the luminosity evolution correction. For future work that is expected to have an improved statistical precision, more detailed knowledge of this correction will be required. \\
\indent To test how sensitive our results are to the assumption that all LRGs are located at the centre of their dark matter haloes, we do two halo model runs where we allow for more flexibility. First, we assume that a fraction of the LRGs is miscentered, following Equation \ref{eq_poff}. We use $p_{\rm off}$ and $R_{\rm off}$ as additional free parameters with a flat uninformative prior in the range [0,1]. The allowed miscentering distribution ranges from the lenses being all correctly centered ($p_{\rm off}$ =$R_{\rm off}$ =0) to all being miscentered and located at the halo scale radius  ($p_{\rm off}$ =$R_{\rm off}$ =1). We find $p_{\rm off}=0.57^{+0.29}_{-0.37}$, $0.26^{+0.39}_{-0.23}$, $0.50^{+0.37}_{-0.36}$ and $0.40^{+0.38}_{-0.30}$ for the low and high redshift bins of LOWZ and CMASS, respectively, with corresponding miscentering radii of $R_{\rm off}=0.37^{+0.41}_{-0.25}$, $0.32^{+0.53}_{-0.29}$, $0.49^{+0.39}_{-0.37}$ and $0.54^{+0.37}_{-0.43}$. The resulting powerlaw parameters are listed in Table \ref{tab_fitpar2}. We find that the powerlaw parameters do not change significantly. The total increase in halo mass of LRGs corresponds to $75^{+38}_{-27}$\%, consistent with our nominal result of $80_{-28}^{+39}$\%. These constraints on the miscentering distribution are in broad agreement with previous galaxy-galaxy lensing and clustering results of CMASS galaxies. For instance, \citet{Miyatake13} report $p_{\rm off}<0.66$ and $R_{\rm off}=0.79^{+0.58}_{-0.38}$, whereas \citet{More14} find $p_{\rm off}=0.34\pm0.18$ and  $R_{\rm off}=2.2^{+1.5}_{-1.3}$.  \\
\indent As a related test, we study how our results change when we assume that a fraction of LRGs are satellites. We use a simple model with only one free parameter, i.e. $M_1$. We follow this approach because we are not interested to determine the satellite HOD (as lensing alone is not very sensitive to that), but because we want to get an estimate how much our results could be affected by ignoring the contribution of satellites. The typical satellite fraction for LRGs is $\sim$10\% and decreases for more massive LRGs \citep{White11,Parejko13,More14}. We therefore put a prior on $M_1$ such that the resulting satellite fractions from our model are between 5\% and 15\%. The resulting luminosity-to-halo mass relation parameters are listed in Table \ref{tab_fitpar2}. The best-fit slopes are consistent with our nominal run, but the amplitude decreases by 1-2$\sigma$. The total increase of halo mass is $82^{+35}_{-29}$\% over the full redshift range of our LRG sample, consistent with our nominal result. The normalisation decreases because the satellites are associated to more massive haloes, with correspondingly larger lensing signals. This lowers the required contribution to the total signal from central LRGs, and hence their mass. This does not happen in the miscentering run, where the lensing signal is merely smeared out, but the integrated signal (and hence the mass) stays the same. \\

\begin{table}
  \caption{Powerlaw parameters, normalisation of the mass-concentration relations and reduced chi-squared values for various runs as described in the text.}   
  \centering
  \begin{tabular}{c c c c c} 
  \hline  \hline
  & & & & \\
  & $M_{0,L}$ & $\beta_L$ & $f_{\rm conc}$ & $\chi^2_{\rm red}$ \\
  & & & & \\
\multicolumn{2}{c}{nominal run} \\
 0.15$<z<$0.29 & $6.43\pm0.52$ & $1.59\pm0.29$ & $0.68^{+0.08}_{-0.06}$ & 1.8 \\
 0.29$<z<$0.43 & $5.14\pm0.44$ & $1.42\pm0.27$ & $0.85^{+0.12}_{-0.11}$ & 1.6 \\
 0.43$<z<$0.55 & $4.08\pm0.51$ & $1.27\pm0.31$ & $0.77^{+0.19}_{-0.15}$ & 1.2 \\
 0.55$<z<$0.70 & $3.74\pm0.60$ & $1.32\pm0.44$ & $0.73^{+0.25}_{-0.20}$ & 1.0 \\
  & & \\
\multicolumn{2}{c}{$e$-correction $\times$ $(1-0.2z)$} \\
 0.15$<z<$0.29 & $5.90\pm0.50$ & $1.56\pm0.29$ &$0.74^{+0.10}_{-0.07}$ & 1.9 \\
 0.29$<z<$0.43 & $5.03\pm0.45$ & $1.44\pm0.27$ &$0.83^{+0.12}_{-0.10}$ & 1.3 \\
 0.43$<z<$0.55 & $3.82\pm0.47$ & $1.19\pm0.35$ &$0.80^{+0.19}_{-0.16}$ & 1.0 \\
 0.55$<z<$0.70 & $3.33\pm0.58$ & $1.43\pm0.53$ &$0.75^{+0.30}_{-0.22}$ & 1.0 \\
  & & \\
\multicolumn{2}{c}{$e$-correction $\times$ $(1+0.2z)$} \\
 0.15$<z<$0.29 & $6.12\pm0.51$ & $1.65\pm0.30$ &$0.75^{+0.10}_{-0.07}$ & 1.9 \\
 0.29$<z<$0.43 & $5.51\pm0.47$ & $1.53\pm0.26$ &$0.84^{+0.12}_{-0.10}$ & 1.4 \\
 0.43$<z<$0.55 & $4.59\pm0.56$ & $1.35\pm0.31$ &$0.83^{+0.19}_{-0.17}$ & 1.3 \\
 0.55$<z<$0.70 & $4.15\pm0.68$ & $1.58\pm0.49$ &$0.73^{+0.25}_{-0.20}$ & 1.0 \\
  & & \\
\multicolumn{2}{c}{miscentering run} \\
 0.15$<z<$0.29 & $6.10\pm0.50$ & $1.61\pm0.29$ & $0.92^{+0.37}_{-0.21}$ & 1.9 \\
 0.29$<z<$0.43 & $5.08\pm0.44$ & $1.41\pm0.27$ & $0.95^{+0.21}_{-0.15}$ & 1.8 \\
 0.43$<z<$0.55 & $3.87\pm0.49$ & $1.26\pm0.31$ & $1.04^{+0.50}_{-0.27}$ & 1.3 \\
 0.55$<z<$0.70 & $3.75\pm0.60$ & $1.31\pm0.44$ & $0.93^{+0.44}_{-0.28}$ & 1.1 \\
  & & \\
\multicolumn{2}{c}{satellite fraction run} \\
 0.15$<z<$0.29 & $4.93\pm0.45$ & $1.39\pm0.26$ & $0.87^{+0.11}_{-0.10}$ & 1.9 \\
 0.29$<z<$0.43 & $4.22\pm0.38$ & $1.27\pm0.25$ & $1.12^{+0.18}_{-0.17}$ & 1.7 \\
 0.43$<z<$0.55 & $3.17\pm0.35$ & $1.02\pm0.29$ & $0.90^{+0.24}_{-0.20}$ & 1.2 \\
 0.55$<z<$0.70 & $2.88\pm0.44$ & $1.08\pm0.43$ & $0.82^{+0.41}_{-0.24}$ & 1.0 \\
  & & \\
  \hline \hline
  \end{tabular}
  \label{tab_fitpar2}
\end{table}


\subsection{Comparison to previous work \label{sec_comp}}
\indent Several previous works have provided mass estimates of LRGs, using gravitational lensing, clustering, abundance matching, or a combination of these. The selection of the samples, the models fit to the data and the definitions of mass, generally differ between these studies, limiting the level of detail with which we can perform a comparison.  \\

\subsubsection{Lensing results}
\indent \citet{Mandelbaum06LRG} measured the masses for a sample of over $4\times10^{4}$ LRGs with spectroscopic redshifts from SDSS-I/II using weak lensing. Bluer and fainter LRGs are discarded using colour-magnitude cuts, as well as LRGs that are likely satellites of larger systems, hence the selection is not identical to ours. The resulting sample is split into a faint and bright part using a cut at $M_r=-22.3$, and the mean luminosity of the two samples is $5.2\times10^{10}$ and $8.6\times10^{10}h^{-2}L_\odot$, respectively. The luminosities are computed in a similar manner as in our work. All LRGs are selected in the redshift range $0.15<z<0.35$ and have a mean effective redshift of 0.24. Masses are estimated using NFW fits plus a baryonic component, where the fitting range is restricted to small scales were the 2-halo term can be neglected. The masses are defined as the enclosed mass in a sphere where the density is 180 times the mean background density, $M_{180b}$, instead of 200 times the mean density which we use; the difference between these definitions is only a few percent. The faint LRGs are found to reside in haloes of masses $2.9\pm0.4 \times10^{13} h^{-1}M_\odot$ and the bright ones in haloes of masses $6.7\pm0.8\times10^{13} h^{-1}M_\odot$. The measurements are shown in Figure \ref{plot_lm2}. We find that the masses of our low-redshift slices are substantially larger than the ones from \citet{Mandelbaum06LRG}. The discrepancy may be caused by the scatter between luminosity and halo mass, which is not included in \citet{Mandelbaum06LRG}. Allowing for a non-zero scatter results in larger halo masses.  \\ 
\indent In \citet{Miyatake13}, the lensing signal of 4,807 CMASS galaxies with $0.47<z<0.59$ is measured in the overlap with CFHTLS using the publicly available CFHTLenS catalogues \citep{Heymans12}. The lensing signal is fit together with the projected clustering signal using a halo model that is similar to the one we have adopted here. Halo masses are defined with respect to 200 times the background density as we do. The average halo mass of CMASS galaxies is found to be $2.3\pm0.1 \times10^{13} h^{-1}M_\odot$. To compare it with our results, we compute the average luminosity of CMASS galaxies with $0.47<z<0.59$ in our catalogue. We find a value of $3.8\times10^{10} h^{-2} L_\odot$ and assume that this number is representative for the average luminosity of the lenses in that work. This estimate is actually a bit too low, as \citet{Miyatake13} also apply a cut on stellar mass to remove the least massive (hence faintest) objects, which we cannot mimic. However, as an indication, we find that if we remove the faintest 10\% of our CMASS sample, the average luminosity only increases to $4.1\times10^{10} h^{-2} L_\odot$, hence it is unlikely that the average luminosity is very far off. We compare the results in Figure \ref{plot_lm2} and find that our CMASS masses are somewhat lower but consistent. \\
\indent Several papers have studied how the average mass of galaxies changes as a function of redshift \citep[e.g.][]{VanUitert11,Choi12,Leauthaud11,Tinker13,Hudson15}, using weak lensing measurements. Since these works do not target LRGs specifically, and since the modelling of the signal differs from our approach, we cannot compare the results in detail. However, both \citet{Tinker13} and \citet{Hudson15} include red, massive galaxies in their work, hence we can at least compare the recovered trends. \\
\indent \citet{Tinker13} use measurements of weak lensing, clustering and the stellar mass function of galaxies in COSMOS to constrain the stellar-to-halo mass relation. Halo masses are defined like ours. The relations they report predict the average $\log_{10}(M_*)$ as a function of halo mass, instead of the average halo mass at a given stellar mass, which is what we measure; these relations are not the same due to intrinsic stellar mass scatter, which is illustrated in their Figure 7. From the right-hand panel of that figure we observe that for passive galaxies, at the average stellar masses of our samples, the mean halo mass is roughly $\sim$0.5 dex lower than what we find. There are many differences between the analyses that could contribute to this difference, such as systematic offsets between stellar mass estimates, the selection of the samples, the definition of mass and the modelling of the signal. Nonetheless, at a stellar mass of $\log_{10}(M_*)\sim$ 11.4 (typical for LRGs), the average halo mass increases from $\sim$$10^{12.9}$ to \mbox{$\sim$$10^{13.2}$ $M_\odot$} between redshifts of 0.88 and 0.36, an increase of almost 100\%, similar to the average growth in halo mass that we find for our LRG sample from redshift 0.62 to 0.21. \\
\indent \citet{Hudson15} use the shape measurements from CFHTLenS to measure the lensing signal for blue and red galaxies in three redshift slices. The lenses are binned in luminosity rather than stellar mass to avoid an Eddington bias due to the larger observational errors on stellar mass compared to luminosity. The stellar mass is then determined using the mean stellar-mass-to-luminosity ratio. For the highest luminosity bin of red lenses, which has a mean stellar mass of $\sim$$2\times10^{11}h_{70}^{-2} M_\odot$, the average halo mass increases from $0.84\pm0.17\times10^{13}h_{70}^{-1}M_\odot$ to $1.32\pm0.33\times10^{13}h_{70}^{-1}M_\odot$ from redshift 0.67 to 0.29. The masses are defined with respect to $\rho_{\rm crit}$ instead of the mean density, resulting in masses that are $\sim$30-40\% smaller compared to ours. Furthermore, the intrinsic scatter between luminosity/stellar mass and halo mass is not accounted for in their modelling, which also leads to smaller masses. Finally, the selection of the lens samples differs. However, the $\sim$58\% increase in average halo mass is comparable to what we find. \\

\subsubsection{Clustering results}
\indent The clustering of LRGs is well-studied in the literature and has been used to derive halo masses \citep[e.g.][]{Blake08,Wake08,Zheng09,Sawangwit11,Nikoloudakis13,Parejko13,Guo14}. \citet{Parejko13} measure the clustering of galaxies with $0.2<z<0.4$ from the LOWZ sample and fit it with a halo model. The probability distribution of halo masses, as shown in their Figure 9, has a mean of $5.2\times10^{13}h^{-1}M_\odot$. We plot it in Figure \ref{plot_lm2} and find that it is larger than our LOWZ measurements. Although not specified, we assume that the mass is defined as $M_{180b}$, as is mentioned in a companion paper \citep{White11}, which is comparable to our definition. Their halo mass distribution is fairly broad, however, and our constraints may well fall inside their 68\% confidence region. \\
\indent \citet{Guo14} measure the clustering of CMASS galaxies divided in three $i$-band magnitude selected samples, which we cannot directly compare to our measurements. They find that going from their faintest to their brightest bins, the peak host halo mass increases from $1.1\times10^{13}h^{-1}M_\odot$ to $3.3\times10^{13}h^{-1}M_\odot$, which is quite comparable to our results. Their masses are defined with respect to the mean density, like ours. However, we measure an `effective' mass and not the peak halo mass, and it is unclear how much these definitions differ. \\
\indent \citet{Zheng09} fit the clustering signal of SDSS-I/II LRGs using an HOD approach. Their LRGs are divided in a faint and bright sample using their $M_g$ magnitude, hence we cannot directly compare. The distribution of halo masses (defined like ours) of these samples peaks at $\sim$$4.5\times10^{13}h^{-1} M_\odot$ and $\sim$$10^{14}h^{-1} M_\odot$, respectively, similar to the values we find for faint and bright LOWZ samples, but note, again, that we measure the effective mass and not the peak halo mass. The scaling of luminosity with host halo mass, which is given by $L_c \propto M^{0.66}$, is consistent with our results. \\
\indent \citet{Wake08} measure the evolution of the clustering signal of galaxies from SDSS and the 2dF-SDSS LRG and QSO Survey \citep[2SLAQ;][]{Cannon06}. They match the selections using colour and magnitude cuts, which complicates the comparison with our results. However, they find that the effective halo masses increase with $\sim$50\% from $z=0.55$ to $z=0.2$, consistent with our findings. \citet{Sawangwit11} study three separate LRGs samples in SDSS, with a mean redshift of 0.35, 0.55 and 0.75. After applying additional selection criteria such that the space density of LRGs is similar to the SDSS-I/II LRG sample, the effective halo mass, defined as the virial mass, is found to be $6.4\pm0.5\times10^{13}h^{-1}M_\odot$, $4.7\pm0.2\times10^{13}h^{-1}M_\odot$ and $4.3\pm0.2\times10^{13}h^{-1}M_\odot$, respectively. We assume that the average luminosity of these three samples is similar to that of the LOWZ sample and show their measurements in Figure \ref{plot_lm2}. We find that their results agree fairly well with ours. \citet{Blake08} and \citet{Nikoloudakis13} study the clustering signal of different samples of LRGs, which is difficult to compare with our results. The typical halo masses of LRGs are found to be a few times $10^{13}h^{-1}M_\odot$, hence broadly in agreement.\\

\subsubsection{Abundance matching results}
\indent \citet{Masaki13} apply an abundance matching technique to N-body simulations in order to construct mock LRG samples whose number density matches that of LRGs in SDSS. From these samples a mock lensing signal is constructed, whose shape agrees well with, but whose amplitude is $\sim$20\% larger than, the measured lensing signal of SDSS LRGs presented in \citet{Mandelbaum13}. They find that their LRGs reside in haloes with a mean virial mass of $5.6\pm0.1\times10^{13}h^{-1}M_\odot$. This definition of mass is $\sim$10\% smaller than $M_{200}$ for masses and concentrations that are typical for LRGs at $z\sim$0.3. We assume the average luminosity equals that of SDSS-I/II LRGs and show the measurement in Figure \ref{plot_lm2}. We find that this mass estimate is in fair agreement with our results.


\subsection{Interpretation}
\begin{figure}
  \resizebox{\hsize}{!}{\includegraphics{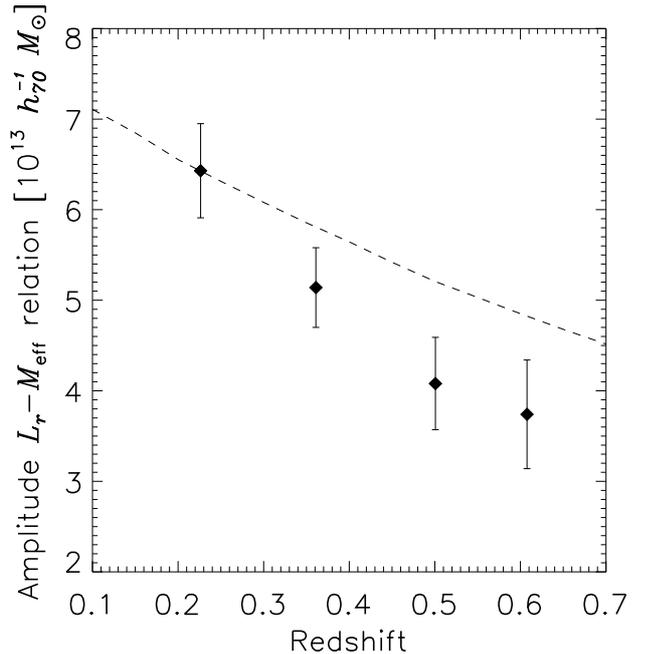}}
  \caption{Evolution of the amplitude of the powerlaw fit between luminosity and halo mass with redshift. The black dashed lines shows the predicted trend from pseudo-evolution \citep{Diemer13} for halo masses that are typical for LRGs, scaled to overlap with our first data point.}
  \label{plot_ampevo}
\end{figure}
\indent We find that $M_{\rm eff}$ of LRGs increases by approximately 80\% from redshift 0.6 to 0.2. This is not only because of dark matter accretion. Part of the growth can be attributed to the so-called pseudo-evolution \citep{Diemer13}. $M_{200}$ is defined with respect to a reference density, which is redshift dependent. Even if a halo is static, i.e. it does not accrete anything,  $M_{200}$ increases towards lower redshift. The halo mass function that we use to determine $M_{\rm eff}$ uses masses that are defined with respect to $\rho_m(z)$. Hence the evolution of the halo mass function is a mix of physical evolution and pseudo-evolution. Therefore, the evolution of $M_{\rm eff}$ is also a mix of the two. \\
\indent Since the halo masses of our LRGs span a relatively narrow range, we can estimate the contribution from pseudo-evolution to our observed increase in mass using the results from \citet{Diemer13}. For halo masses that are typical for our LRGs, $M_{200}$ increases by approximately 33\% from $z=0.6$ to $z=0.2$ due to pseudo-evolution, as is illustrated in Figure \ref{plot_ampevo}. If we fit the amplitude of the pseudo-evolution curve to our measurements, we find that $\chi^2_{\rm red}=1.3$, providing weak support that the slope is steeper and requires additional dark matter accretion. \\
\indent It is interesting to compare our growth rates to those obtained from simulations. Ideally, we would like to compare to hydrodynamical simulations, but those are not available for the mass range we are interested in. However, a comparison with dark-matter only simulations is interesting as well, as a good/poor agreement points towards the relevance of baryonic physics. We first compare to the two Millennium simulations, for which growth and merger rates have been derived in \citet{Fakhouri10}. They find that the growth rate is well described by:
\begin{eqnarray}
\dot{\langle M \rangle}_{\rm mean}=46.1M_\odot \rm{yr}^{-1}\left( \frac{\mathit{M}}{10^{12}\mathit{M}_\odot}\right)^{1.1}  \nonumber \\
\times(1+1.11z)\sqrt{\Omega_{\rm M}(1+z)^3+\Omega_{\Lambda}}.
\end{eqnarray}
Using this equation, we find that a halo of mass $10^{13}M_\odot$ grows by 38\% from redshift 0.6 to 0.22, whilst a halo of mass $5\times10^{13}M_\odot$ grows by 43\%. More recently, \citet{Wetzel14} use hydrodynamical simulations to measure the amount of physical accretion (i.e.  after accounting for pseudo-evolution) for haloes with masses $10^{11}-10^{13}M_\odot$. However, for masses in the range $10^{13}-10^{14}M_\odot$, which are typical for LRGs, they use dark-matter only simulations and derive a growth of $\sim$10\% between $z$=0.5 to $z$=0 at scales smaller than a few hundred kpc. Together with pseudo-evolution, the total growth amounts to $\sim$43\%, comparable to the results of \citet{Fakhouri10}. \\
\indent \citet{Wetzel14} only study isolated haloes; LRGs cluster strongly, so the typical halo in that work may not be very representative for LRGs. However, no environment selection cuts are made in \citet{Fakhouri10} and the derived growth rates are similar, hence this seems to be unimportant. In fact, \citet{Fakhouri10b} measure the growth rate as a function of environment and find a weak trend of a decreasing growth rate towards denser environments, but the statistics for massive haloes is poor due to the low number of haloes. \\
\indent We find that the growth rates predicted from dark-matter only simulations are marginally consistent with our results. Our measurements suggest a larger growth, particularly towards more massive haloes. This could point toward the impact of baryons; particularly, AGN feedback may have an impact on the distribution of matter in massive haloes \citep[][]{Duffy10,Velliscig14} and their accretion history. However, we need to improve the statistics of our measurements before we can make stronger claims. \\
\indent Our conclusion about the evolution of the luminosity-to-halo mass relation depends, however, on the validity of the assumption of pure passive evolution. The absence of a clear tilt in Figure \ref{plot_magdist} supports this view. Nonetheless, some residual star formation and/or mergers may also affect the luminosities, which could have an impact on our conclusions.  \\
\indent The amount of star formation that is allowed in LRGs is limited, based on their colour evolution \citep{Wake06,Maraston09}. Also, spectral analyses from SDSS-I/II LRGs and CMASS galaxies point toward a very low fraction of galaxies that either form stars or have AGN activity \citep{Greisel13,Steel13}. There are indications that intermediate mass early-type galaxies have a low level of ongoing star formation \citep[e.g][]{Kaviraj07,Schawinsky07,Salim10}, but it seems unlikely that it is sufficiently strong to affect our conclusions.\\
\indent Mergers also complicate a purely passive luminosity evolution scenario. The merging history of LRGs or massive early-type galaxies has received considerable attention in recent years \citep[e.g][]{Tal12,Lopez12,Gabor12,Bedorf13,Ruiz14}. We focus here on the results from \citet{Tojeiro10}, as they estimate the amount of luminosity growth in LRGs due to mergers. Their analysis is based on the measured luminosity function and clustering strength of LRGs in the SDSS in the redshift range $0.15<z<0.5$, from which they deduce that the average luminosity of LRGs increases by 1.5-6\% Gyr$^{-1}$ due to mergers, depending on luminosity, such that the growth mainly happens for the faintest LRGs. For LRGs with $M_{r,0.1}<-22.8$, the evolution is consistent with passive evolution. In Table 4 of that work, luminosity growth rates from recent works are compared, suggesting that their values are fairly representative. \\
\indent \citet{Tojeiro10} propose two type of mergers that could contribute to the luminosity growth: a merger between an LRG and a small companion, or a merger between two small companions, whose combined luminosity is sufficient to classify it as LRG. In the first scenario, the luminosities of LRGs increase towards lower redshifts; hence they are overestimated compared to the luminosities of LRGs that have evolved through pure passive evolution. \\
\indent To estimate how that may impact our derived luminosity-to-halo mass relations and its evolution, we assume that the luminosities of our three LRGs samples increase through mergers with 2, 4 and 6\% Gyr$^{-1}$, from the bright to the faint bin, respectively, without affecting the halo masses, i.e. we assume that the mass-to-light ratio of the smaller companions equals zero. This provides us with an upper limit on the bias in the derived halo mass growth. We use the highest redshift slice as our reference, and lower the luminosities of the lower redshift slices, to mimic how the evolution would have looked in the absence of mergers. For example, between $z=0.6$ and $z=0.2$, approximately 3.3 Gyr passed, so we lower the luminosity of the L1z1 bin by a factor $1.06^{3.3}\approx 1.2$. We apply this factor to the average passively evolved luminosities, i.e. we do not apply it to the individual luminosities before the selection of the samples. After this adjustment, we fit the luminosity-to-halo mass relation again. We find that the retrieved slopes of the luminosity-to-halo mass relation are within the error bars of our nominal results. The amplitudes move up by $\sim$2$\sigma$ for our $0.15<z<0.29$ redshift slice, and by $\sim$1$\sigma$ for our $0.29<z<0.43$ slice. The average growth in halo mass becomes $116_{-35}^{+50}$\%, consistent with our nominal result. We ignored here that mergers also cause growth in mass, hence the LRGs move diagonally rather than horizontally in the luminosity-to-halo mass plane. If the mass-to-light ratio of the smaller companions is similar to that of the LRG, they move along the luminosity-to-halo mass relation and no bias is caused; if this ratio is larger than that of LRGs, the actual growth in halo mass would be smaller than our nominal value. \\
\indent The second channel for luminosity growth through mergers described by Tojeiro is through the merging of two faint galaxies, which are individually not bright enough to be selected as an LRG. As the formation history differs from that of the typical LRG that formed at high redshift without much activity afterwards, the properties of their haloes may be different, complicating the interpretation of the measured trends. \\
\indent Apart from that, there are several other physical processes that complicate the interpretation. For example, star formation and mergers may be linked in LRGs \citep{Kaviraj11}. In addition, some high-redshift LRGs may not become low-redshift LRGs, as they may be tidally disrupted or merge with other galaxies. Furthermore, AGN activity could trigger star formation, leading to too blue colours to match the LRG colour selection at lower redshift. The subset of LRGs whose luminosity-to-halo mass evolution is most easy to interpret is probably the brightest one: \citet{Tojeiro12} find that the brightest CMASS galaxies are most likely to evolve in SDSS-I/II LRGs, and \citet{Tojeiro10} derive that the brightest LRGs experience the least amount of luminosity growth from mergers. \\
\indent To summarise the above: although passive evolution accounts for the bulk of the luminosity evolution of LRGs, there are several processes that complicate this picture. This affects our ability to select and compare LRGs and their progenitors at different redshifts. In order to extract the full information content on the evolution of LRGs that is contained in our measurements, one should compare with numerical simulations that contain all relevant physical processes. No luminosity evolution correction needs to be applied in that case, since there is no need to match the samples at different redshifts, as long as the selection can be matched in observations and simulations. The range in luminosity and redshift of the LRG sample studied in this work would form an ideal observational test case for such a study. \\


\section{Mass-concentration relation \label{sec_massconc}}
\begin{figure}
  \resizebox{\hsize}{!}{\includegraphics{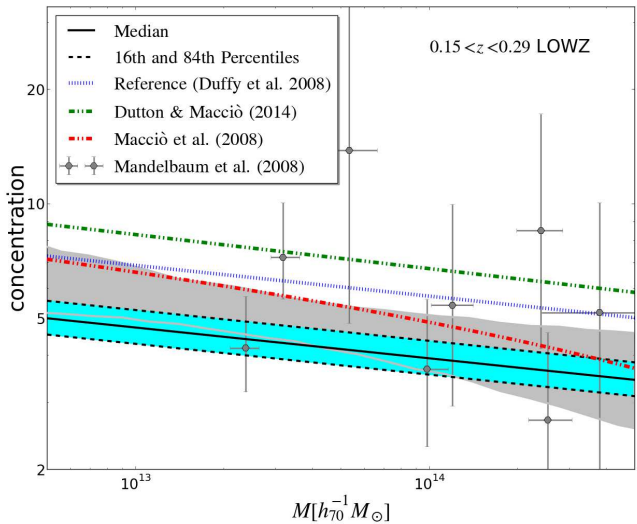}}
  \caption{Best-fit mass-concentration relation derived for the $0.15<z<0.29$ LOWZ bin. The solid black line indicates the best-fit result and the turquoise area the 68\% confidence intervals. The blue dotted line shows the reference relation from \citet{Duffy08}, whose amplitude we fit in the halo model. The red and green dot-dashed lines show the relations from \citet{Maccio08} and \citet{Dutton14}, respectively. The black points and the gray area are the results from direct fits to lensing data from \citet{Mandelbaum08}.}
  \label{plot_massconc}
\end{figure}
\begin{figure}
  \resizebox{\hsize}{!}{\includegraphics{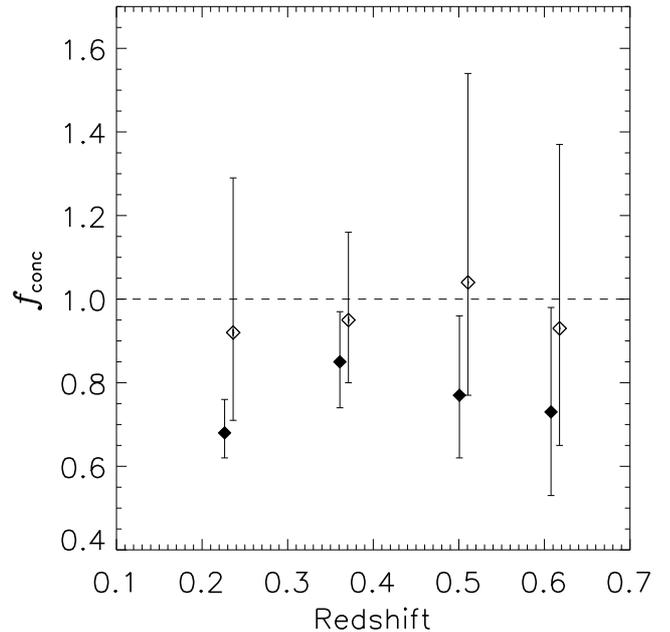}}
  \caption{Best-fit normalisation of the mass-concentration relation from \citet{Duffy08} for the four redshift slices. Filled symbols show the results for the nominal run, open symbols for the miscentering run. The open symbols have been shifted to the right for improved visibility.}
  \label{plot_fconc}
\end{figure}
\indent As discussed in Section \ref{sec_hm}, we assume a functional form for the mass-concentration relation, i.e. the one from \citet{Duffy08}, but we allow the overall normalisation of this relation to vary in the fit. The resulting constraints for the $0.15<z<0.29$ LOWZ bins are shown in Figure \ref{plot_massconc}, together with the nominal \citet{Duffy08} model. The constraints on $f_{\rm conc}$ as a function of redshift are shown in Figure \ref{plot_fconc} and listed in Table \ref{tab_fit2}. \\
\indent We find that the derived mass-concentration relations are lower than the reference model. The largest discrepancy is for our lowest redshift slice, where we find \mbox{$f_{\rm conc}=0.68^{+0.08}_{-0.06}$}. For the other redshift slices, our results are about 1$\sigma$ below the relation from \citet{Duffy08}. There is no evidence for a trend of $f_{\rm conc}$ with redshift, although the errors of the higher redshift bins are still fairly large. It does suggest that the redshift scaling of this relation is well captured by its functional form (Equation \ref{eq_mc}).\\
\indent In Figure \ref{plot_massconc}, we also plot the relation from \citet{Maccio08}, based on dark matter only simulations using the WMAP3 cosmology \citep{Hinshaw09}, and the relation from \citet{Dutton14}, who derived the mass-concentration relation with N-body simulations using the Planck cosmology \citep{Planck14}. All the relations are shown for a redshift of 0.23. These relations derived from dark matter only simulations consistently predict higher concentrations in the mass range that we study.  \\
\indent It is possible that the low normalisation results from choices in our modelling. Therefore, we also derive the normalisation of the mass-concentration relation for the halo model runs where we either allow a fraction of the LRGs to be miscentered, where we allow a non-zero satellite fraction, and for the alternative luminosity evolution correction runs. The results are listed in Table \ref{tab_fitpar2}. For the alternative luminosity evolution run, the constraints on $f_{\rm conc}$ do not change significantly. For the other two runs, the normalisations increase and become consistent with unity, as is shown in Figure \ref{plot_fconc}. By adding either satellites or a miscentered component, we basically shift power from small scales to intermediate scales. Consequently, to fit the same data, the lensing signal of the central LRGs needs to become steeper, i.e. the concentrations need to increase. Note that the errors on $f_{\rm conc}$ increase as well, due to the increased freedom in the model (particularly for the miscentering run).\\
\indent In Figure \ref{plot_massconc} we also show the results from \citet{Mandelbaum08}, who derived the mass-concentration relation by combining lensing measurements for L$*$ type galaxies, galaxy groups traced by LRGs and the maxBCG cluster sample in the SDSS. The model used for their main results did not include a miscentering component, but they used a minimum scale in the fit of \mbox{500 $h^{-1}$kpc} for the maxBCG sample to reduce the impact of miscentering. We only show the measurements that overlap with our range of halo masses. We find that our results are consistent with theirs. Towards lower masses, the mass-concentration relation of \citet{Mandelbaum08} is in agreement with theoretical expectations, but at larger masses, the normalisation is lower, as can been seen in their Figure 5. As their lenses span a broad range of mass, they can fit a more flexible mass-concentration relation to their measurements, resulting in a broader confidence region compared to ours. \\
\indent The mass-concentration relation of CMASS galaxies is also derived in \citet{Miyatake13}. Their results are very similar to ours: using their fiducial model, in which all galaxies are located at the centre of dark matter haloes, they find a normalisation of $f_{\rm conc}=0.78\pm0.12$ with respect to the relation of \citet{Maccio07}. However, when they include a miscentering component identical to ours, the normalisation factor becomes consistent with unity. \\
\indent The results from \citet{Miyatake13} and the ones presented here suggest that the low normalisation of the mass-concentration relation can be explained by allowing for satellites and/or a miscentering component in the modelling, although the constraints on the miscentering parameters are still very weak. To understand whether the effect is real, one either needs to move towards larger data-sets whose corresponding statistical errors are much smaller, or alternatively, one can make use of group catalogues with spectroscopically identified members, to study the brightest group members and the satellites separately. The overlap between the Galaxy And Mass Assembly \citep[GAMA;][]{Driver09,Driver11,Robotham11} and the Kilo Degree Survey (KiDS; Kuijken et al., in prep) could offer an excellent dataset for such a work.\\
\indent However, it is interesting to consider alternative explanations, such as baryonic processes that can lower the concentrations. AGN feedback, for example, may affect the distribution of dark matter at the relevant radii, as has been studied in \citet{Velliscig14}. In that work, several baryonic processes such as cooling, supernova feedback and AGN feedback are implemented in hydrodynamical simulations to study their effect on the density profiles of haloes. For halo masses typical for LRGs, AGN feedback is the dominant process and affects the dark matter distribution out to several times the virial radius, such that the density is lower at small scales. A similar conclusion was reached in \citet{Duffy10}; whilst cooling leads to more concentrated haloes, this effect is counteracted by AGN feedback, which could cause the concentrations to fall $\sim$15\% below the ones from dark matter only simulations. 
 

\section{Conclusion \label{sec_conc}}
We study the evolution of the luminosity-to-halo mass relation of LRGs, combining SDSS photometry and spectroscopy with the excellent imaging data from the RCS2, which enables us to measure the weak lensing signal up to $z$$\sim$0.6. We use stellar population synthesis modelling to compute the correction to account for passive luminosity, which is thought to dominate the luminosity evolution of LRGs, enabling us to compare low-redshift LRGs with their predecessors at higher redshift. We split the LOWZ and CMASS galaxies in two redshift slices and three luminosity bins each, resulting in twelve lens samples, and fit a halo model to the lensing signals of the three luminosity bins simultaneously, for each redshift slice separately. The halo mass estimates that we obtain are broadly consistent with various literature results that are based on a variety of measurement techniques, but span a considerably larger combined range of luminosity and redshift. \\
\indent We find a typical value of $\sim$1.4 for the slope of the luminosity-to-halo mass relation and no evidence that it changes with redshift. The amplitude of this relation, however, does increase significantly with redshift. We find that the average halo mass of LOWZ galaxies increases by $25_{-14}^{+16}$\% from $\langle z\rangle=0.36$ to $\langle z\rangle=0.22$. The halo masses of CMASS galaxies grows by $10_{-20}^{+25}$\% from $\langle z\rangle=0.6$ to $\langle z\rangle=0.5$. If CMASS galaxies are the predecessors of the LOWZ ones, the total growth of LRGs from $\langle z\rangle=0.6$ to $\langle z\rangle=0.22$ is $80_{-28}^{+39}$\%. \\
\indent This growth in halo mass is somewhat larger than what is expected for pure pseudo-evolution, i.e. the evolution in the definition of $M_{200}$ caused by the change of the mean background density as the universe expands, which by itself causes an apparent halo mass growth of $\sim$33\% at typical LRG masses between $z$=0.6 and $z$=0.2. Our measurements provide weak support for additional dark matter accretion.  \\
\indent We have tested the sensitivity of these results against changes in the luminosity evolution and changes in the halo model. We find that the inferred slopes of the luminosity-to-halo mass relation do not significantly change. The amplitude of this relation decreases by at most 2$\sigma$, in the case where we include a satellite component in the halo model. The inferred average growth in halo mass of LRGs does not change by more than 1$\sigma$. Hence for this work, systematic errors are likely subdominant to the statistical ones. For future work that uses measurements with a higher statistical precision, advances in the modelling of the luminosity evolution and of the set up of the halo model, are necessary to avoid biases in the results. Such measurements will be highly valuable to constrain the impact of baryonic processes on the distribution of dark matter, which is essential for a correct and optimized exploitation of future cosmic shear surveys such as Euclid \citep{Laureijs11}. \\
\indent We also constrain the overall normalisation of the mass-concentration relation for each of the four redshift slices. We find that for our lowest redshift slice, the best-fit relation is lower than what is expected from dark matter only simulations. However, if we allow for miscentering or for the contribution from satellites in the halo model, the normalisation increases and becomes consistent with the results from dark matter-only simulations. \\

\paragraph{Acknowledgements \\ \\} 
We would like to thank Alexie Leauthaud and Claudia Maraston for useful discussions and suggestions. This work was supported by a grant from the German Space Agency DLR. HH, MC and RH acknowledge support from NWO VIDI grant number 639.042.814 and ERC FP7 grant 279396.
\\
\indent This work is based on observations obtained with MegaPrime/MegaCam, a joint project of CFHT and CEA/DAPNIA, at the Canada-France-Hawaii Telescope (CFHT) which is operated by the National Research Council (NRC) of Canada, the Institute National des Sciences de l'Univers of the Centre National de la Recherche Scientifique of France, and the University of Hawaii. We used the facilities of the Canadian Astronomy Data Centre operated by the NRC with the support of the Canadian Space Agency. \\
\indent Funding for SDSS-III has been provided by the Alfred P. Sloan Foundation, the Participating Institutions, the National Science Foundation, and the U.S. Department of Energy Office of Science. The SDSS-III web site is http://www.sdss3.org/.\\
\indent SDSS-III is managed by the Astrophysical Research Consortium for the Participating Institutions of the SDSS-III Collaboration including the University of Arizona, the Brazilian Participation Group, Brookhaven National Laboratory, University of Cambridge, Carnegie Mellon University, University of Florida, the French Participation Group, the German Participation Group, Harvard University, the Instituto de Astrofisica de Canarias, the Michigan State/Notre Dame/JINA Participation Group, Johns Hopkins University, Lawrence Berkeley National Laboratory, Max Planck Institute for Astrophysics, Max Planck Institute for Extraterrestrial Physics, New Mexico State University, New York University, Ohio State University, Pennsylvania State University, University of Portsmouth, Princeton University, the Spanish Participation Group, University of Tokyo, University of Utah, Vanderbilt University, University of Virginia, University of Washington, and Yale University. 
\bibliographystyle{aa}

\begin{appendix}

\section{Luminosity evolution scatter \label{sec_ap_int}}
The evolution corrected luminosities of galaxies become increasingly uncertain with redshift. This is mainly due to the intrinsic variation in properties of the LRGs, as illustrated in the bottom panel of Figure \ref{plot_kecorr}, where the difference between the $e$-correction curves increases with redshift for the different models. \\
\indent To obtain a rough estimate of how intrinsic scatter in the ($k+e$)-corrected absolute magnitudes affects our results, we do a simple test. We start with randomly drawing $10^5$ redshifts and ($k+e$)-corrected absolute magnitudes from the original distribution, i.e. the one that is shown in the bottom panel of Figure \ref{plot_magdist}. We assume that these magnitudes are the intrinsic ones. Next, we assign a mass to each object using our nominal best-fit luminosity-to-halo mass relation from Table \ref{tab_fitpar2}. Note that our conclusions do not sensitively depend on the choice of slope and offset. We compute an NFW profile for each object at 100  logarithmically spaced radial bins between 0.05 h$_{70}^{-1}$ Mpc and 1 \mbox{h$_{70}^{-1}$ Mpc}. We stack the NFW profiles of objects that fall inside a lens bin as defined in Figure \ref{plot_magdist}, and fit an NFW to the resulting profile using the mean redshift of those objects.\\
\indent Next, we simulate the intrinsic scatter in absolute magnitude by assuming that it can be described by a Gaussian whose width increases as $\sigma=0.3z$. This particular choice covers the range of $e$-correction curves of the SSP models that are in reasonable agreement with the colours of the LRGs, as shown in Figure \ref{plot_kecorr}. We draw a random number from this Gaussian and add that to the `intrinsic' magnitudes: these are the `observed' magnitudes. We stack the NFW profiles, but now using the `observed' magnitudes to select the lenses. Again, we fit an NFW profile to the stacked profile using the mean redshift of the observed objects. The impact of intrinsic luminosity scatter is then estimated from the ratio of this `observed' mass to the intrinsic one. We find that the mass typically changes by a few percent. The largest difference is found for the L3z4 bin, where the difference is $\sim$10\%. However, in all cases, the error is considerably smaller than the statistical errors. We therefore conclude that intrinsic scatter of the luminosities can be safely ignored. 

\section{Systematic tests \label{sec_ap_sys}}
\begin{figure}
  \resizebox{\hsize}{!}{\includegraphics[angle=-90]{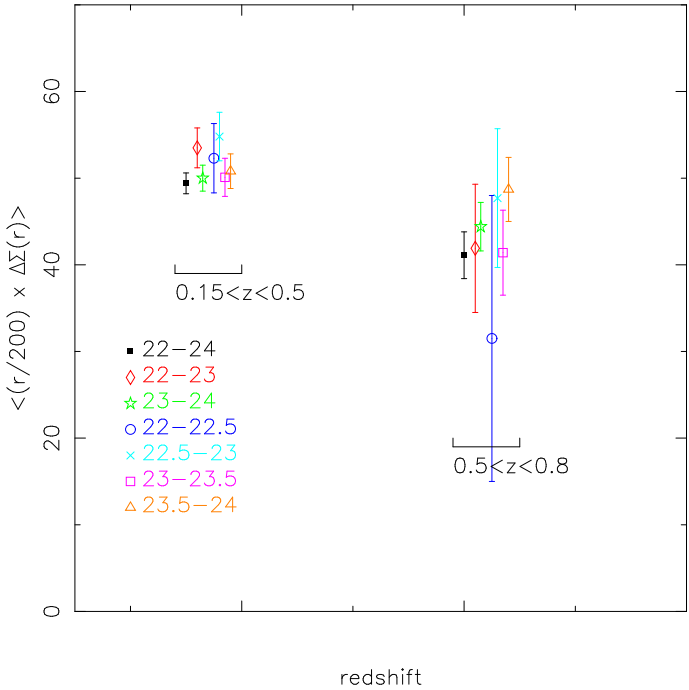}}
  \caption{The weighted mean of the lensing signal times the radius for a low-z and high-z LRG lens sample, measured with different source samples as indicated in the plot. For both lens samples, we find consistent results, suggesting that our measurement pipeline is robust and does not introduce a redshift dependent bias.} 
  \label{plot_test}
\end{figure}
Since this work aims at measuring the redshift dependence of the lensing signal, we have to be particularly careful with systematic effects that mimic a redshift scaling. One issue that requires attention is the computation of the mean lensing efficiency. For example, if we miss the high redshift tail of the source redshift distribution, the mean lensing efficiency would be biased increasingly low for higher lens redshifts, which would mimic a redshift dependence in the lensing signal. Motivated by the differences in mean lensing efficiency computed from the photometric redshift catalogue from \citet{Ilbert09} and \citet{Ilbert13} at high lens redshifts, we perform the following test.\\
\indent We divide our source sample in six different samples, with magnitude cuts of $22<m_{r'}<23$, $23<m_{r'}<24$, and $22<m_{r'}<22.5$, $22.5<m_{r'}<23$, $23<m_{r'}<23.5$, $23.5<m_{r'}<24$.  We split the LRGs in a low-redshift sample with $0.15<z<0.5$ and a high-redshift one with $0.5<z<0.8$, and measure their lensing signals using each source sample. Note that for each source sample, we separately determine the shear signal around random points, the mean lensing efficiencies using the \citet{Ilbert13} photometric redshift catalogues, the contamination of source galaxies around the lenses, and the noise bias correction. We measure the weighted mean of the lensing signal times the projected separation, as that is roughly a constant with radius, over the range 0.15 $<r<$ 10 h$_{70}^{-1}$ Mpc. We show these values for the two LRG samples in Figure \ref{plot_test}. \\
\indent We find that the weighted mean of the lensing signal of each lens sample is consistent for the various source samples. Clearly, the measurements are correlated for the source samples that overlap in apparent magnitude, but the four bins of 0.5 magnitude width are more or less independent (not completely as they are similarly affected by cosmic variance). This result strongly suggests that the measurement process is robust. \\
\indent We have repeated this test using the \citet{Ilbert09} photometric redshift catalogue instead. For the \mbox{$0.15<z<0.5$} LRG sample the results are consistent, but for the high-z one, some bins differ by 2-3$\sigma$. This points at a problem with the source redshift distribution used to compute the lensing efficiencies. Since the \citet{Ilbert13} photometric redshift catalogue gives consistent results for the different source samples, even at high redshifts, this suggests that the source redshift distribution is sufficiently accurately determined with the latter, but not with the former \citep[see also the discussion in][]{Hoekstra15} .

\section{Fitting Methodology \label{ap_mcmc}}
We use Bayesian inference techniques to determine the posterior probability distribution $P(\lambda | \calD)$ of the model parameters $\lambda$, given the data $\calD$.  According to Bayes' theorem,
\begin{equation}
P(\lambda | \calD)= {P(\calD | \lambda) \, P(\lambda)
\over P(\calD)} \,,
\end{equation}
where $P(D | \lambda)$ is the likelihood of the data given the model parameters, $P(\lambda)$ is the prior probability of these parameters, and
\begin{equation}
P(\calD)= \int P(\calD | \lambda) \, P(\lambda) \,
\rmd\lambda\,,
\end{equation}
is called the evidence. Since we do not intend to perform model selection, the evidence just acts as a normalisation constant which needs not to be calculated. Therefore, the posterior distribution $P(\lambda|D)$ is given by
\begin{equation}
P(\lambda | \calD) \propto \exp\left[\frac{-\chi^2(\lambda)}{2} \right]\,,
\end{equation}
where
\begin{equation}\label{eq:chi2}
\chi^2(\lambda)=\chi^2_{\rm ESD} = \sum_{k=1}^{N_{\rm lum}}
\sum_{j=1}^{N_{\Delta\Sigma}}
\left[\frac{\Delta\Sigma(R_j|L_k) - \widetilde{\Delta\Sigma}(R_j|L_k)}
{\sigma_{\Delta\Sigma}(R_j|L_k)}\right]^2\,.
\end{equation}
$\widetilde{\Delta\Sigma}$ denotes the model prediction, $\sigma_{\Delta\Sigma}$ is the corresponding error, $N_{\rm lum} = 3$ and $N_{\Delta\Sigma}=12,12,12,11$ for the low and high redshift bins of the LOWZ and CMASS samples, respectively. For our fiducial model, the set of model parameters is $\lambda_{\rm fid} = (M_{\rm mean},\sigma_{\rm \log M},f_{\rm conc})$, where:
\begin{itemize}
\item $M_{\rm mean}$ is the mean halo mass for the $k$-th luminosity bin;
\item $\sigma_{\rm \log M}$ is the scatter of $\langle N_\rmc|M \rangle$;
\item $f_{\rm conc}$ is the normalisation of the $c(M)$ relation.
\end{itemize}
When exploring model variations (see Section \ref{sec_sens}) we employ $\lambda_{\rm miscen} = (\lambda_{\rm fid}, p_{\rm off},{R}_{\rm off})$  and $\lambda_{\rm satfrac} = (\lambda_{\rm fid}, M_1)$. For all model parameters, we adopt a flat, sufficiently wide prior such that the results are not biased. \\
\indent We sample the posterior distribution of our model parameters given the data using a Monte-Carlo Markov chain (MCMC). In particular, we implement the Metropolis-Hastings algorithm to construct the MCMC.  At any point in the chain, a trial model is generated using a method specified below. The $\chi^2$ statistic for the trial model, $\chi^2_{\rm try}$, is calculated using Equation (\ref{eq:chi2}). This trial model is accepted to be a member of the chain with a probability given by
\begin{equation}
P_{\rm accept}=\left\{ \begin{array}{ll}
        1.0 & \mbox{if $\chi^2_{\rm try} \le \chi^2_{\rm cur}$} \\
  {\rm exp}[-(\chi^2_{\rm try}-\chi^2_{\rm cur})/2] &
              \mbox{if $\chi^2_{\rm try} > \chi^2_{\rm cur}$}
\end{array}\right. \,,
\end{equation}
where $\chi^2_{\rm cur}$ denotes the $\chi^2$ for the current model in the chain. We initialize the chain from a random position in our multi-dimensional parameter space and obtain a chain of $\sim$500,000 models. We discard the first $10,000$ models (the burn-in period) allowing the chain to sample from a more probable part of the distribution. We use this chain of models to estimate the confidence levels on the parameters and on the lensing signal, as shown in Figure \ref{plot_gg2}. \\
\indent A proper choice of the proposal distribution is very important in order to achieve fast convergence and a reasonable acceptance rate for the trial models. The posterior distribution in a multi-dimensional parameter space, such as the one we are dealing with, will have degeneracies and in general can be very difficult to sample from. We have adopted the following strategy to overcome these difficulties. During the first half of the burn-in stage, we chose an independent Gaussian proposal distribution for every model parameter, as is common for the Metropolis-Hastings algorithm.  Half-way through the burn-in stage, we perform a Fisher information matrix analysis at the best-fit model found thus far. The Fisher information matrix, given by
\begin{equation}
F_{ij} = -\frac{\partial^2 \ln\calL}{\partial \lambda_i \partial
    \lambda_j}\,,
\end{equation}
is a $N_\rmp \times N_\rmp$ symmetric matrix, where $N_\rmp$ denotes the number of parameters in our model, and $\calL \propto e^{-\chi^2/2}$ is the likelihood. The inverse of the Fisher matrix gives the covariance matrix, $\bC$, of the posterior constraints on the model parameters. More importantly, the eigenvectors of the covariance matrix are an excellent guide to the degeneracies in the posterior distribution, and the corresponding eigenvalues set a scale for how wide the posterior ought to be in a given direction. Therefore, for the second half of the burn-in period, we utilize this information and use a proposal distribution which is a multi-variate Gaussian centered at the current value of the parameters and with a covariance given by the aforementioned matrix. In practice, the trial model
($\lambda_{\rm
  try}$) can be generated from the current model ($\lambda_{\rm
  cur}$) using
\begin{equation}
\lambda_{\rm try}=\lambda_{\rm cur} + \zeta\,\bA \bx \,,
\end{equation}
where $\bx$ is a vector consisting of $N_\rmp$ standard normal deviates, the matrix $\bA$ is such that $\bA\bA^T=\bC$, and $\zeta$ is a parameter that we have chosen to achieve an average acceptance rate of $\sim$30\%.  We repeat the Fisher matrix analysis once again at the end of the burn-in period (using the best-fit model found thus far) and use the covariance matrix to define our proposal distribution to be used for the MCMC. This strategy has proven to be extremely efficient in sampling  posterior distributions for similar studies \citep[see e.g.][]{Cacciato13,Cacciato14}.

\section{k-corrected only results}
\begin{figure*}[t!]
  \resizebox{\hsize}{!}{\includegraphics{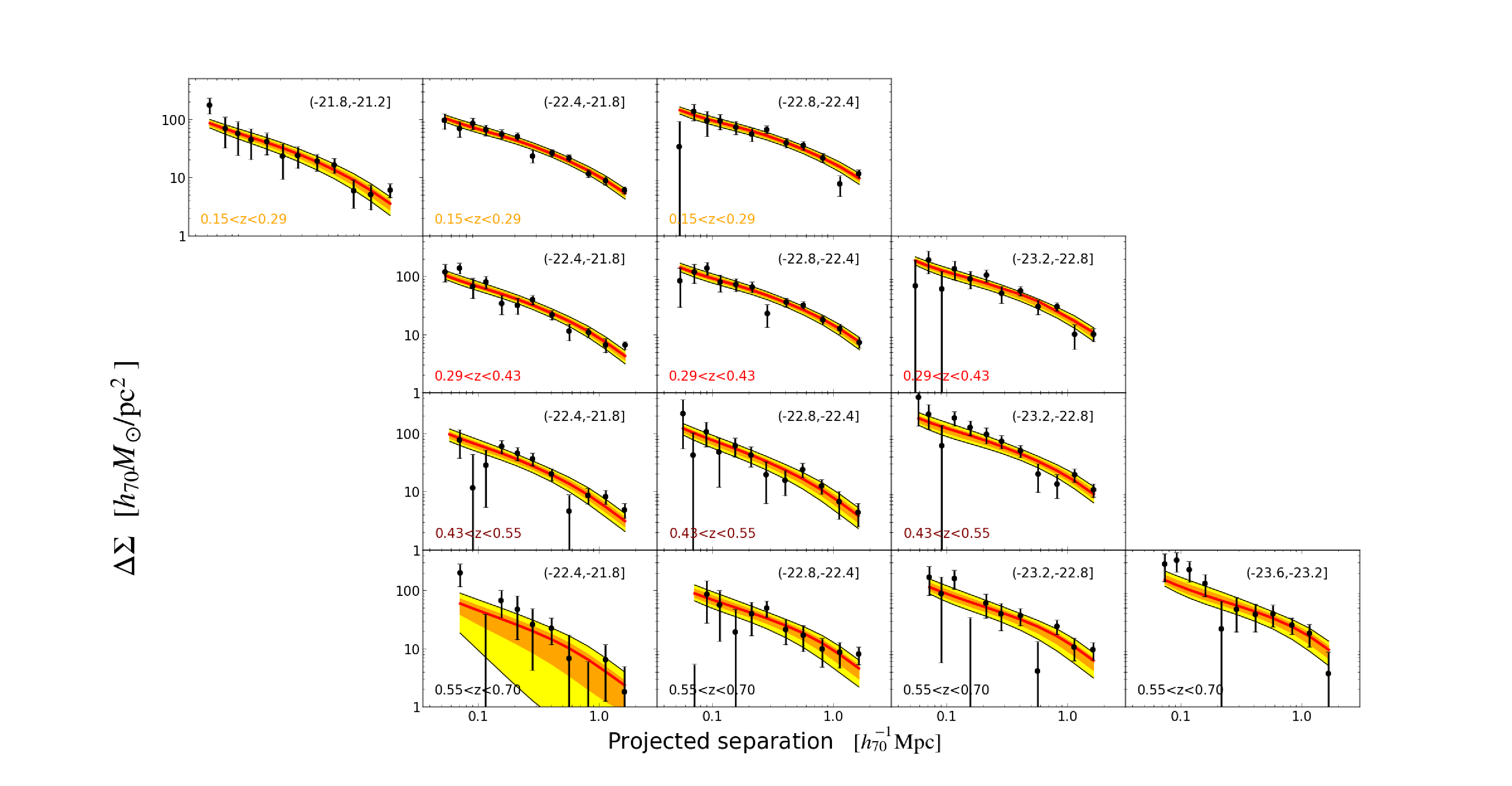}}
  \caption{The lensing signal $\Delta \Sigma$ of LOWZ (top two rows) and CMASS (bottom two rows) lenses as a function of projected separation for the four luminosity bins (after the $k$-correction is applied). The solid red lines show the best-fit halo model, the orange and yellow regions the 1 and 2$\sigma$ model uncertainty, respectively. We fit the signal on scales between 0.05 and 2 $h^{-1}_{70}$ Mpc.}
  \label{plot_gg}
\end{figure*}

In this work, we have attempted to account for the evolution of the luminosities of LRGs by applying a luminosity evolution correction. Unfortunately, this correction is uncertain and will also add scatter to the corrected luminosities as LRGs cover a range of intrinsic properties such as formation age and metallicity, which complicates the interpretation of the results. Therefore, it is also interesting to apply only the $k$-correction to the absolute magnitudes. The resulting luminosities are closer to the real luminosities, i.e. they have smaller scatter. Interpreting any trend in the luminosity-to-halo mass relation will be harder as both the luminosity and the halo mass may evolve simultaneously, but comparison to for example simulations should be more straightforward. \\
\indent Hence we divide the lens sample into bins of $k$-corrected absolute magnitude and redshift. Details of the lens samples can be found in Table \ref{tab_fit}. The selection is also illustrated in Figure \ref{plot_magdist}. For each subsample, we stack the lensing signal of all the lenses inside that bin and show that in Figure \ref{plot_gg}. We fit the halo model using the same set-up as before and show the best-fit models, together with the model uncertainties, in Figure \ref{plot_gg}. The corresponding effective masses and normalisations of the mass-concentration relation can be found in Table \ref{tab_fit}. \\
\begin{table}
\renewcommand{\tabcolsep}{0.12cm}
  \caption{Properties of the lens bins (after $k$-correction)}   
  \centering
  \begin{tabular}{c c c c c c c c c c} 
  \hline  \hline
  & & & & & & & \\
 & M$_r$ & N$_{\rm lens}$ & $\langle z \rangle$ & $\langle L_r \rangle$ & $M_{\rm eff}$ & $f_{\rm conc}$ \\
 & (1) & (2) & (3) & (4) & (5) & (6) & \\
  & & & & & & \\
  \hline
  & & & & & & \\
\multicolumn{3}{c}{0.15$<$$z$$<$0.29 (LOWZ)} \\
L1z1 & [-21.8,-21.2] & 1110  & 0.204 & 0.68 & $2.60^{+0.57}_{-0.53}$ & \\  
L2z1 & [-22.4,-21.8] & 3784  & 0.225 & 1.04 & $4.48^{+0.56}_{-0.53}$ & $0.78^{+0.09}_{-0.09}$ \\  
L3z1 & [-22.8,-22.4] & 954  & 0.234 & 1.59 & $8.73^{+1.17}_{-1.08}$ & \vspace{2mm} \\ 
\multicolumn{3}{c}{0.29$<$$z$$<$0.43 (LOWZ)} \\
L2z2 & [-22.4,-21.8] & 4725 & 0.351 & 1.10 & $3.17^{+0.50}_{-0.46}$ &  \\  
L3z2 & [-22.8,-22.4] & 2901 & 0.368 & 1.62 & $5.78^{+0.83}_{-0.76}$ & $0.85^{+0.12}_{-0.12}$ \\  
L4z2 & [-23.2,-22.8] & 981  & 0.374 & 2.31 & $8.64^{+1.38}_{-1.25}$ &  \vspace{2mm} \\  
\multicolumn{3}{c}{0.43$<$$z$$<$0.55 (CMASS)} \\
L2z3 & [-22.4,-21.8] & 7798 & 0.496 & 1.09 & $2.10^{+0.46}_{-0.41}$ & \\  
L3z3 & [-22.8,-22.4] & 4197 & 0.506 & 1.62 & $2.48^{+0.62}_{-0.54}$ & $0.85^{+0.19}_{-0.17}$ \\  
L4z3 & [-23.2,-22.8] & 1646 & 0.506 & 2.31 & $7.18^{+1.34}_{-1.21}$ &  \vspace{2mm}\\  
\multicolumn{3}{c}{0.55$<$$z$$<$0.70 (CMASS)} \\
L2z4 & [-22.4,-21.8] & 2532  & 0.584 & 1.17 & $1.08^{+0.78}_{-0.72}$ & \\  
L3z4 & [-22.8,-22.4] & 4357  & 0.599 & 1.67  & $2.79^{+0.89}_{-0.77}$ & $0.69^{+0.27}_{-0.19}$ \\  
L4z4 & [-23.2,-22.8] & 3365  & 0.615 & 2.35 & $3.97^{+1.15}_{-0.99}$ & \\  
L5z4 & [-23.6,-23.2] & 1314  & 0.625 & 3.37 & $6.98^{+2.02}_{-1.74}$ & \\  
  & & & & & & \\
  \hline \hline
 &&\\
  \end{tabular}
  \tablefoot{(1) absolute magnitude range (after $k$-correction); (2) number of lenses; (3) mean redshift; (4) mean luminosity \mbox{[10$^{11} h_{70}^{-2} L_\odot$]} (after $k$-correction); (5) best-fit halo mass \mbox{[10$^{13} h_{70}^{-1} M_\odot$]}; (6) best-fit normalisation of the mass-concentration relation.}
  \label{tab_fit}
\end{table} 

\end{appendix}

\end{document}